\documentclass[a4paper,11pt]{article}
\pdfoutput=1 % if your are submitting a pdflatex (i.e. if you have images in pdf, png or jpg format)
\usepackage{jheppub}	% jheppub includes hyperref, color, natbib, amsmath, amssymb, epsfig, graphicx
\usepackage[utf8]{inputenc}
\usepackage[english]{babel}
\usepackage[dvipsnames]{xcolor}
\usepackage{makeidx}
\usepackage{amsfonts}
\usepackage{mathrsfs}
\usepackage{tensor}
\usepackage[autostyle]{csquotes}
\usepackage{subfig}
\usepackage[export]{adjustbox}
\usepackage{adjustbox}
\newcommand{\ft}[2]{{\textstyle\frac{#1}{#2}}}

\newdimen\tableauside\tableauside=1.0ex
\newdimen\tableaurule\tableaurule=0.4pt
\newdimen\tableaustep
\def\phantomhrule#1{\hbox{\vbox to0pt{\hrule height\tableaurule
width#1\vss}}}
\def\phantomvrule#1{\vbox{\hbox to0pt{\vrule width\tableaurule
height#1\hss}}}
\def\sqr{\vbox{%
  \phantomhrule\tableaustep
\hbox{\phantomvrule\tableaustep\kern\tableaustep\phantomvrule\tableaustep}%
  \hbox{\vbox{\phantomhrule\tableauside}\kern-\tableaurule}}}
\def\squares#1{\hbox{\count0=#1\noindent\loop\sqr
  \advance\count0 by-1 \ifnum\count0>0\repeat}}
\def\tableau#1{\vcenter{\offinterlineskip
  \tableaustep=\tableauside\advance\tableaustep by-\tableaurule
  \kern\normallineskip\hbox
    {\kern\normallineskip\vbox
      {\gettableau#1 0 }%
     \kern\normallineskip\kern\tableaurule}%
  \kern\normallineskip\kern\tableaurule}}
\def\gettableau#1 {\ifnum#1=0\let\next=\null\else
  \squares{#1}\let\next=\gettableau\fi\next}
\def\be{\begin{equation}}
\def\ee{\end{equation}}
\def\bea{\begin{eqnarray}}
\def\eea{\end{eqnarray}}
\newcommand{\nn}{\nonumber}

\title{The dark side of fuzzball geometries}
\author[a,b]{M. Bianchi,}
\author[a,b]{D. Consoli,}
\author[a]{A. Grillo,}
\author[b]{J.F. Morales}
\affiliation[a]{Dipartimento  di  Fisica,  Universit\`a  di  Roma  ``Tor  Vergata'', Via della Ricerca Scientifica, 00133 Roma, Italy}
\affiliation[b]{I.N.F.N.  Sezione  di  Roma  ``Tor  Vergata'', Via della Ricerca Scientifica, 00133 Roma, Italy}
\emailAdd{massimo.bianchi@roma2.infn.it}
\emailAdd{dario.consoli@univie.ac.at}
\emailAdd{alfredo.grillo@roma2.infn.it}
\emailAdd{morales@roma2.infn.it}
\abstract{Black holes absorb any particle impinging with an impact parameter below a critical value. We show that 2- and 3-charge fuzzball geometries exhibit a similar trapping behaviour for a selected choice of the impact parameter of  incoming massless particles.  This suggests that the blackness property of black holes arises as a collective effect whereby each micro-state absorbs a specific channel.}

\preprint{PREPRINT}
 \clearpage

\begin{document}

\maketitle
\flushbottom

\section{Introduction}

The interest in objects colloquially known as black holes (BH's) has been revived not only by their role in the generation of the first gravitational wave signal detected by the LIGO-Virgo collaboration \cite{Abbott:2016blz} but also by the possibility that primordial BH's may account for a (small) fraction of the dark matter in the universe \cite{Clesse:2017bsw} and rotating BH's and similar objects may accelerate cosmic rays thanks to Penrose mechanism \cite{Penrose:1971uk, Bianchi:wip}.

In String Theory it is natural to describe BH's as ensembles of micro-states represented by smooth, horizonless geometries without closed time-like curves, the so called ``fuzzballs"  \cite{Lunin:2001jy,Lunin:2002iz, Mathur:2003hj,Lunin:2004uu,Mathur:2005zp, Skenderis:2008qn,Mathur:2008nj}.  The counting of micro-states for extremal 3- and 4-charged black hole states in five and four dimensions has proven to be very successful \cite{Strominger:1996sh, Breckenridge:1996sn,Maldacena:1996ky, Maldacena:1997de, Maldacena:1998bw}, while the identification of the corresponding geometries in the supergravity regime has revealed to be much harder \cite{Giusto:2004id, Giusto:2004ip,Bena:2005va, Berglund:2005vb, Saxena:2005uk,Bena:2006kb,Bena:2007kg,Bena:2007qc,Giusto:2009qq,Giusto:2011fy,Lunin:2012gp,Giusto:2013bda,Gibbons:2013tqa,Bena:2015bea,Lunin:2015hma,Bena:2016agb, Bena:2016ypk,Pieri:2016cqz, Bianchi:2016bgx, Pieri:2016pdt, Bianchi:2017bxl}. To go one step further one can probe fuzzball geometries with particles, waves and strings and test the proposal at the dynamical level \cite{Bianchi:2017sds}.  Elaborating on our recent work on 2-charge systems \cite{Bianchi:2017sds}, our present focus will be on the geodetic motion of massless particles on a class of 3-charge micro-state geometries introduced in \cite{Bena:2017xbt}. This should capture the relevant physics not only for large impact parameters where the eikonal approximation of the scattering process is valid even quantitatively \cite{Amati:1987wq,Amati:1987uf, Amati:1988tn,Amati:1990xe,Hashimoto:1996kf,Hashimoto:1996bf, Garousi:1996ad,DAppollonio:2010krb,Bianchi:2011se, DAppollonio:2013mgj, DAppollonio:2013okd,DAppollonio:2015oag}, but also for small impact parameters whereby the particles get trapped or absorbed, at least at a qualitative level. We leave the analysis of waves and strings or other classes of smooth geometries (such as JMaRT \cite{Jejjala:2005yu}) to the future. 

The picture that emerges from our analysis is that the blackness property of black holes arises as a collective/statistical effect where each micro-state absorbs a specific channel. More interestingly, this universal property of fuzzball geometries suggests the possible existence of more exotic distributions of micro-state geometries looking effectively as gravitational filters obscuring only a band in the light spectrum of distant sources, or more bizarre black looking objects such as rings, spherical shells, etc. A more detailed analysis should take into account radiation damping, {\it i.e.} the energy lost in gravitational wave emission by an accelerated particle. Contrary to the case of an accelerated charged particle, we expect gravitational brems-strahlung to be anyway negligible for a vast range of kinematical parameters.   

The paper is organised as follows. In section \ref{geomotfuzzgeo} we introduce the class of micro-state geometries we will consider, discuss the general behaviour of massless geodesics 
in these backgrounds and summarise our results. In particular we will introduce the notions of {\it turning points} and {\it critical geodesics}, characterising geodesics that  either bounce back to infinity or get trapped spinning around the gravitational source, respectively. In Section \ref{BHnofuzz}-\ref{3chargefuzz}  we analyse the behavior of  massless geodesics in the  case of 3-charge black holes, 2-charge and 3-charge fuzzballs respectively.  The analysis of 2-charge fuzzballs is performed in full generality, while  the analysis of the 3-charge  case  is restricted  to  geodesic motion along or perpendicular to the plane of the string profile characterising
the fuzzball. The latter case lacks spherical symmetry and exhibits an intricate non-completely separable dynamics. A simple solution in this class is presented in some detail. 
%In Appendix \ref{notsotoy},   a not-so-toy model  of the general case is studied perturbatively in the limit where the breaking of rotational symmetry of the background is small.
Section \ref{outlook} contains our conclusions and outlook.

%%%%%%%%%%%%%%%%%%%%%%%%%%%%%%%%%%%%%%%%%%%%%%%%%%%%%%%%%%%%%%%%%%%%%%%%%%%%%%%%%
%%%%%%%%%%%%%%%%%%%%%%%%%%%%%%%%%%%%%%%%%%%%%%%%%%%%%%%%%%%%%%%%%%%%%%%%%%%%%%%%%
%%%%%%%%%%%%%%%%%%%%%%%%%%%%%%%%%%%%%%%%%%%%%%%%%%%%%%%%%%%%%%%%%%%%%%%%%%%%%%%%%
%%%%%%%%%%%%%%%%%%%%%%%%%%%%%%%%%%%%%%%%%%%%%%%%%%%%%%%%%%%%%%%%%%%%%%%%%%%%%%%%%
%%%%%%%%%%%%%%%%%%%%%%%%%%%%%%%%%%%%%%%%%%%%%%%%%%%%%%%%%%%%%%%%%%%%%%%%%%%%%%%%%
%%%%%%%%%%%%%%%%%%%%%%%%%%%%%%%%%%%%%%%%%%%%%%%%%%%%%%%%%%%%%%%%%%%%%%%%%%%%%%%%%

\section{Overview and summary of results}
\label{geomotfuzzgeo}

In this section we introduce the  fuzzball geometries we will be interested in and summarise our results. We write down the general form of the metric, the Lagrangian governing the dynamics of massless neutral particles and the geodesic equations. We then identify the conjugate momenta and the Hamiltonian and describe how to take advantage of the isometries when present. We also discuss the classification of the geodesics when the system is integrable.

\subsection{The 3-charge fuzzball  metrics}

We will consider 3-charge BPS micro-state geometries belonging to the general class constructed in\footnote{In the notation of this reference, we focus on solutions with $k=1$, $m=0$ and $n$ an arbitrary positive integer.} \cite{Bena:2017xbt}. The ten-dimensional metric can be written in the form
\be
ds^2={\sqrt{ Z_1 Z_2} \over   Z^2  } \, ds_6^2 + \sqrt{\frac{Z_1}{Z_2}}ds_{\mathcal{T}_4}^2.   \label{ds10}
\ee
where $ds_{\mathcal{T}_4}^2$ is the metric on a $T^4$ torus (or a K3 surface, in fact) while the 6-dimensional metric $ds_6^2$ describes a 5-dimensional space-time times a compact circle of radius $R_y$. This manifold can be parametrized with coordinates $\{t,\vec{{X}},y\}$ or alternatively by introducing the null coordinates $u=\frac{t-y}{\sqrt{2}}$ and $v=\frac{t+y}{\sqrt{2}}$ and the oblate spheroidal coordinate system
\bea
{X}_1+{\rm i} {X}_2=\sqrt{\rho^2+a^2} \,\sin {\vartheta}\, e^{ {\rm i} \varphi}  \qquad, \qquad     {X}_3+{\rm i} {X}_4=\rho \, \cos {\vartheta}\, e^{ {\rm i} \psi} \quad .
\eea
By doing so one obtains
\be
\begin{aligned}
\label{metricQ1Q5Qp}
&
ds_6^2=g_{mn} dx^m dx^n=- 2 \left(dv+\beta_m dx^m \right)\left( du  +  \gamma_m dx^m \right) +  Z^2 \, ds_4^2\,.
\end{aligned}
\ee
where $ds_4^2$ is the flat metric of $\mathbb{R}^4$
 \be
\begin{aligned}
\label{metricQ1Q5Qpds4}
ds_4^2=\left(\rho^2{+}a^2\cos^2 {{\vartheta}}\right)\left(\frac{d\rho^2}{\rho^2{+}a^2}+d{\vartheta}^2\right)+\left(\rho^2+a^2\right)\sin^2 {{\vartheta}} \,  d\varphi^2 + \rho^2 \cos^2 {\vartheta}\, d\psi^2 \, .
\end{aligned}
\ee
The functions $Z_1$, $Z_2$, $Z$, $\beta_m$,  $\gamma_m$\footnote{For the class of solutions we are interested in the components $\beta_\rho$, $\beta_\vartheta$, $\beta_u$, $\beta_v$ and $\gamma_u$ are identically zero.} depend on the coordinates $\vec{{x}}$ of $\mathbb{R}^4$ and on $v$, their explicit expression is as follows
\bea
Z_1 & =&1+\frac{L^2_1}{\rho^2+a^2c_{\vartheta}^2 }+ \frac{\,{\varepsilon}_1R^2 \, \Delta_n\,s_{\vartheta}^2\, \cos 2\phi  }{   L_5^2(\rho^2+a^2c_{\vartheta}^2) } 
\qquad\qquad~~ \nn
Z_2 = 1+\frac{L^2_5}{\rho^2+a^2c_{\vartheta}^2}  
\\
Z_4^2 &=&{ 2\, {\varepsilon}_4^2 R^2  \Delta_n \, s_{\vartheta}^2\, \cos^2 \phi \over (\rho^2+a^2 c_{\vartheta}^2)^2 }      
\qquad \qquad\qquad \qquad\qquad\quad~
Z^2 = Z_1 Z_2-Z_4^2 \nn
\\
\beta_\varphi &=& \frac{ a^2 R \,s_{\vartheta}^2  }{  \rho^2+a^2c_{\vartheta}^2} \qquad\qquad\qquad\qquad  ~~~~~~
\beta_\psi = -\frac{ a^2 R \,c_{\vartheta}^2  }{  \rho^2+a^2c_{\vartheta}^2}      \nn \\
\gamma_\varphi &=& \alpha\, \beta_\varphi   -  {n\, {\varepsilon}_1\,R\over 2 L_5^2}\,\Delta_n \cos 2\phi\, s_{\vartheta}^2  
\qquad~
\gamma_\psi=-\alpha \,\beta_\psi      
\qquad\qquad~~~~~ 
\gamma_v={{\mathcal{F}}_n}   \nn
\\
\gamma_{\vartheta} &=& -\frac{{\varepsilon}_1 R}{2 L_5^2} \, \Delta_n\, \sin 2\phi\, s_{\vartheta} \,c_{\vartheta}  
\qquad \qquad ~~
\gamma_\rho = - { {\varepsilon}_1 R \over 2L_5^2} \, {\Delta_n \over \rho}\, \sin 2\phi\, s_{\vartheta}^2 \label{solution1}
\eea
%%%%%%%%%%%%%%%%%%%%%%%%%%%%%%%%%%%%%%%%%%%%%%%%%%%%%%%%%%%%%%%%%%%%%%%%%%%%%%%%%
with $s_{\vartheta}= \sin{\vartheta}, c_{\vartheta}= \cos{\vartheta}$ and 
\be
\begin{aligned}
  \phi  &=\varphi + \frac{n v}{R }    \qquad      R={  R_y \over \sqrt{2} }   \\
  {\mathcal{F}}_n  &= -\frac{{\varepsilon}_4^2}{2a^2}\left[1-\left(\frac{\rho^{2}}{\rho^2+a^2}\right)^n\right] \\ 
 \Delta_n  &=  \frac{a^2}{\rho^2+a^2 }\left(\frac{\rho^2}{\rho^2+a^2}\right)^n         
\\
\alpha  &= 1 - {\cal F}_n-{ n\,  {\varepsilon}_1 \over 2 L_5^2}\,\Delta_n \, \cos 2\phi\, s_{\vartheta}^2\label{solution2}
\end{aligned}
\ee
Regularity of the metric near $\rho = 0$, ${\vartheta} = {\pi/2}$  requires \cite{Bena:2017xbt}
\be
\label{regularitycond}
a^2 =  \frac{L^2_1L^2_5}{2 R^2} - \frac{{\varepsilon}_4^2}{2} \quad , \quad
{{\varepsilon}_4^2={\varepsilon}_1\left(1+\frac{a^2n}{L^2_5}\right)} \,.
\ee
  The conserved charges and the angular momenta $J$ and $\tilde{J}$ are given by
\be
Q_1 = L_1^2 \quad , \quad
Q_5 = L_5^2 \quad , \quad
Q_P=\frac{{\varepsilon}_4^2n}{2} \quad , \quad
J=\tilde{J}=\frac{R a^2}{ \sqrt{2}}\neq 0 \,.
\ee
or equivalently
\be
J_\varphi = J + \tilde{J} = \sqrt{2} R a^2
\qquad
,
\qquad
J_\psi = J - \tilde{J} = 0
\ee
 We will study the scattering of massless neutral particles in the following special cases of the  family of BPS metrics introduced above:
  
  \begin{itemize}
  
  \item{3-charge non-rotating black holes: Recovered as the $a\to 0$ limit of the 3-charge metric.}
  
  \item{2-charge fuzzball: Obtained  by setting $\varepsilon_1 =n = 0$  in the 3-charge metric.}
  
   \item{3-charge fuzzball:  The general case restricted to the planes $\vartheta = 0$ and $\vartheta = \pi/2$.  }
  
  \end{itemize}

\subsection{The geodesics}

% \be
%\begin{aligned}
%\label{metricQ1Q5Qpds4}
%ds_4^2=\left(\rho^2+a^2c_{{\vartheta}}^2\right)\left(\frac{d\rho^2}{\rho^2+a^2}+d{\vartheta}^2\right)+\left(\rho^2+a^2\right)s_{{\vartheta}}^2   d\varphi^2 + \rho^2c_{\vartheta}^2d\psi^2 \, .
%\end{aligned}
%\ee
      
%    \be
%    \phi=\varphi+{n v\over p\, R}
%    \ee
%    with $n$, $p$ integers and $R=R_y/\sqrt{2}$  is the radius of the  direction 
%   

We are interested in null geodesics in the 6-dimensional geometry that solve the Euler-Lagrange equations derived from the Lagrangian
 \be
\begin{aligned}
\label{masslessequivmetric}
\mathcal{L} &= \ft12 g_{mn}\, \dot x^m \dot x^n  \, ,
\end{aligned}
\ee
with $g_{mn}$ the six-dimensional metric,\footnote{The never vanishing factor $\frac{\sqrt{Z_1 Z_2}}{Z^2}$ in front of the 6-dimensional metric (\ref{metricQ1Q5Qp}) can be absorbed in a redefinition of the affine parameter $\tau$, and neglected when dealing with 6-dimensional geodesics.}
 and dots denoting derivatives with respect to an affine parameter $\tau$. Null geodesics are specified by solutions $x^m(\tau)$ of the Euler-Lagrange equations satisfying  ${\cal L}=0$.
Equivalently one can introduce the Hamiltonian
\be
{\cal H} = P_m \dot x^m-{\cal L} = \ft12 g^{mn}\, P_m P_n  
\ee
expressed in terms of the conjugate momenta   
\be
P_m={\partial {\cal L} \over  \partial \dot x^m} =g_{mn}\, \dot x^n \,. \label{pdotx} 
\ee
It will prove useful to keep in mind that
%\footnote{Indeed
%  $P_u = -\frac{P_y + E}{\sqrt{2}} \leq 0$ and $   P_v = \frac{P_y - E}{\sqrt{2}} \leq 0 $. 
%}
\be
2P_u P_v = E^2 - P_y^2 \geq 0 \, ,
\ee
where $E$ and $P_y$ are the momenta conjugate to $t$ and $y$, respectively.
In the Hamiltonian formulation, geodesics are described by the velocities
\be
\dot x^m={\partial {\cal H} \over \partial P_m}
\ee 
 with $P_m$ a  solution of the system of equations\footnote{We notice that the equations of motion  imply
\be
 \dot{\cal H}=   g^{mn}\, P_m \left(\dot{P}_n+ {\partial {\cal H} \over \partial x^n} \right)=0
\ee
 so, one of the equations of motion, let us say the one for $\rho$  can be replaced by ${\cal H}=0$. 
}
\bea
 2 {\cal H}&=&   g^{mn}\, P_m P_n  =0 \\
  \dot P_m &=& -{\partial  {\cal H}  \over \partial x^m}  
\eea
  The metric is independent of the variables $u$ and $\psi$, so the momenta $P_u$ and $P_\psi$ will always be conserved.
The Hamiltonian can be written in the compact form  
  \bea
  {\cal H}  &=&    -    \, P_u\, \widehat{P}_v   +{1\over 2Z^2} \left[  \frac{(\rho^2 + a^2) \widehat P_\rho^2 }{\rho^2+a^2c_{\vartheta}^2}
+\frac{\widehat P_{\vartheta}^2 }{\rho^2+a^2c_{\vartheta}^2} + \frac{\widehat P_\varphi^2   }{(\rho^2 + a^2)s_{\vartheta}^2}
+\frac{\widehat  P_\psi^2  }{ \rho^2 \, c_{\vartheta}^2}   \right]   \label{hh}
\eea 
  in terms of the shifted momenta
  \be
  \widehat P_m
  %=P_m -\gamma_m P_u -\beta_m (P_v -\gamma_v  P_u ) 
  = P_m -\beta_m (P_v -\gamma_v P_u)-\gamma_m P_u
  \ee
  The velocities become
  \bea
  \label{xdotPhat}
  \dot{\rho} &=&  \frac{(\rho^2 + a^2) \widehat P_\rho }{ Z^2 (\rho^2+a^2c_{\vartheta}^2)} \qquad , \qquad
 \dot{\vartheta} = \frac{\widehat P_{\vartheta} }{Z^2(\rho^2+a^2c_{\vartheta}^2)} \nn\\
 \dot{\varphi} &=&  \frac{\widehat P_\varphi   }{ Z^2 (\rho^2 + a^2)s_{\vartheta}^2}\qquad , \qquad
 \dot{\psi} = \frac{\widehat  P_\psi }{  Z^2\, \rho^2 \, c_{\vartheta}^2}    
  \eea
   with more involved formulae for $\dot{u}$ and $\dot{v}$. 
  The Hamiltonian constraint ${\cal H}=0$ can be solved by taking
  \be
   \widehat P_\rho =\pm \left( { \rho^2+a^2 c_{\vartheta}^2 \over \rho^2+a^2}  \right)^{1\over 2} \left[   2Z^2\,  P_u\, \widehat{P}_v   - \frac{\widehat P_{\vartheta}^2 }{\rho^2+a^2c_{\vartheta}^2} - \frac{\widehat P_\varphi^2   }{(\rho^2 + a^2)s_{\vartheta}^2}
-\frac{\widehat  P_\psi^2  }{ \rho^2 \, c_{\vartheta}^2}  \right]^{1\over 2}   \label{widerho}
  \ee
   with minus and plus signs for the branches along which the particle approaches or leaves the gravitational target, respectively. We notice that according to (\ref{xdotPhat}) 
  $ \widehat P_\rho$ determines the radial velocity of the particle.  
Starting from infinity, $\rho(\tau)$ monotonously decreases until it reaches a point $\rho_*$ where $\widehat{P}_\rho$ vanishes and flips sign. This is said to be an {\it inversion} (or {\it turning}) point. 
Since $\rho$ is a monotonous function along this branch it can  be used {\it in principle}  to parametrize the evolution time, expressing all remaining coordinates $x^m(\rho)$ as a function of $\rho$ instead of the affine parameter $\tau$. In practice, this is possible only when the system is integrable. Examples of integrable geodesics occur for BH's with or without angular momenta, 2-charge circular fuzzballs and geodesics along the plane orthogonal to the string profile in the 3-charge system. The most difficult and interesting case (motion along the plane of the profile in the 3-charge case) eludes this simplistic analysis and will be addressed in section \ref{3chpi2}.

\begin{center}
\begin{figure}[t]
 %\subfloat[][$b > b_c$]
{\includegraphics[scale=0.28]{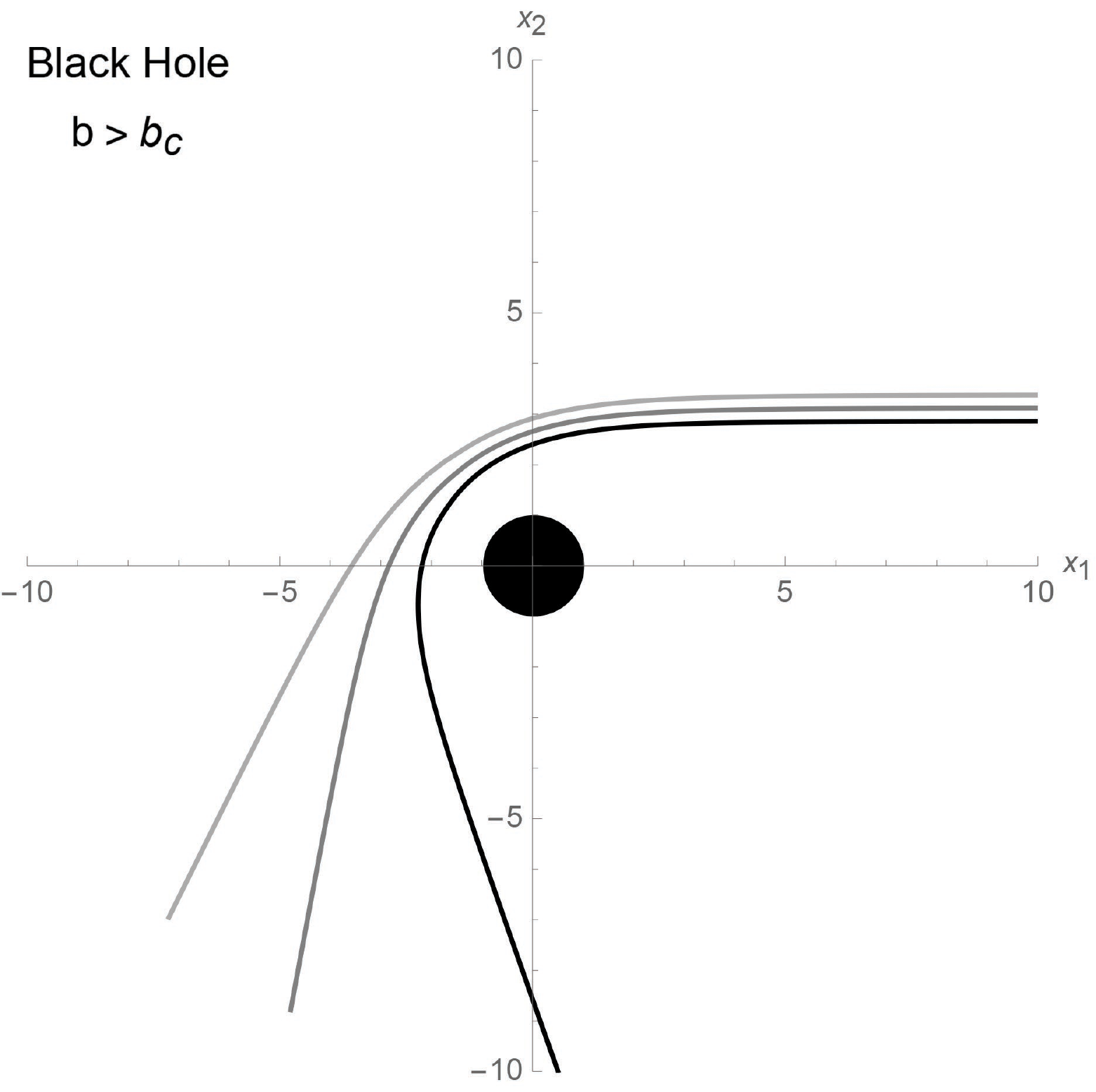} }
\quad
%\subfloat[][$b = b_c$]
{\includegraphics[scale=0.28]{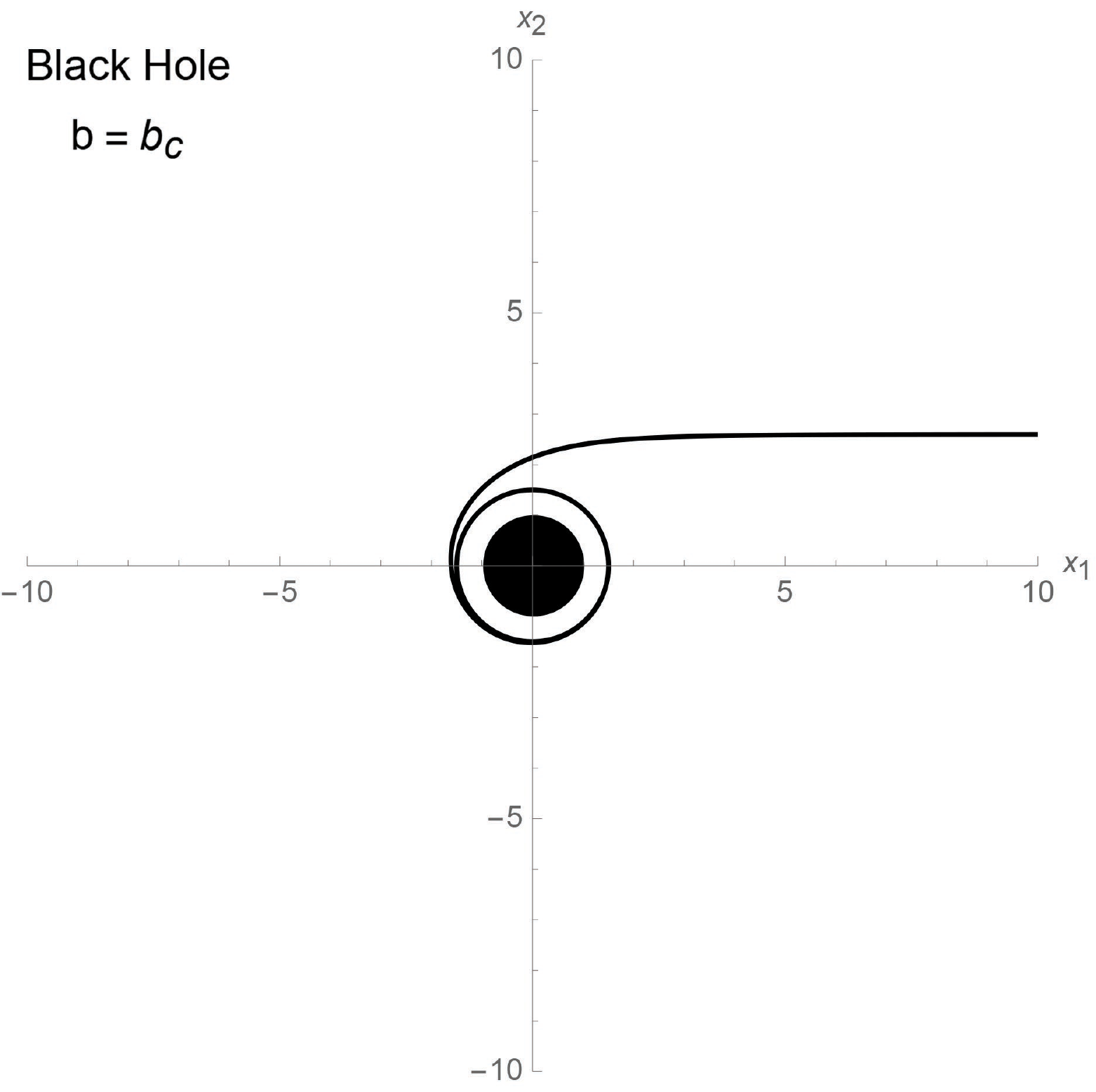} }
\quad
%\subfloat[][$b < b_c$]
{\includegraphics[scale=0.28]{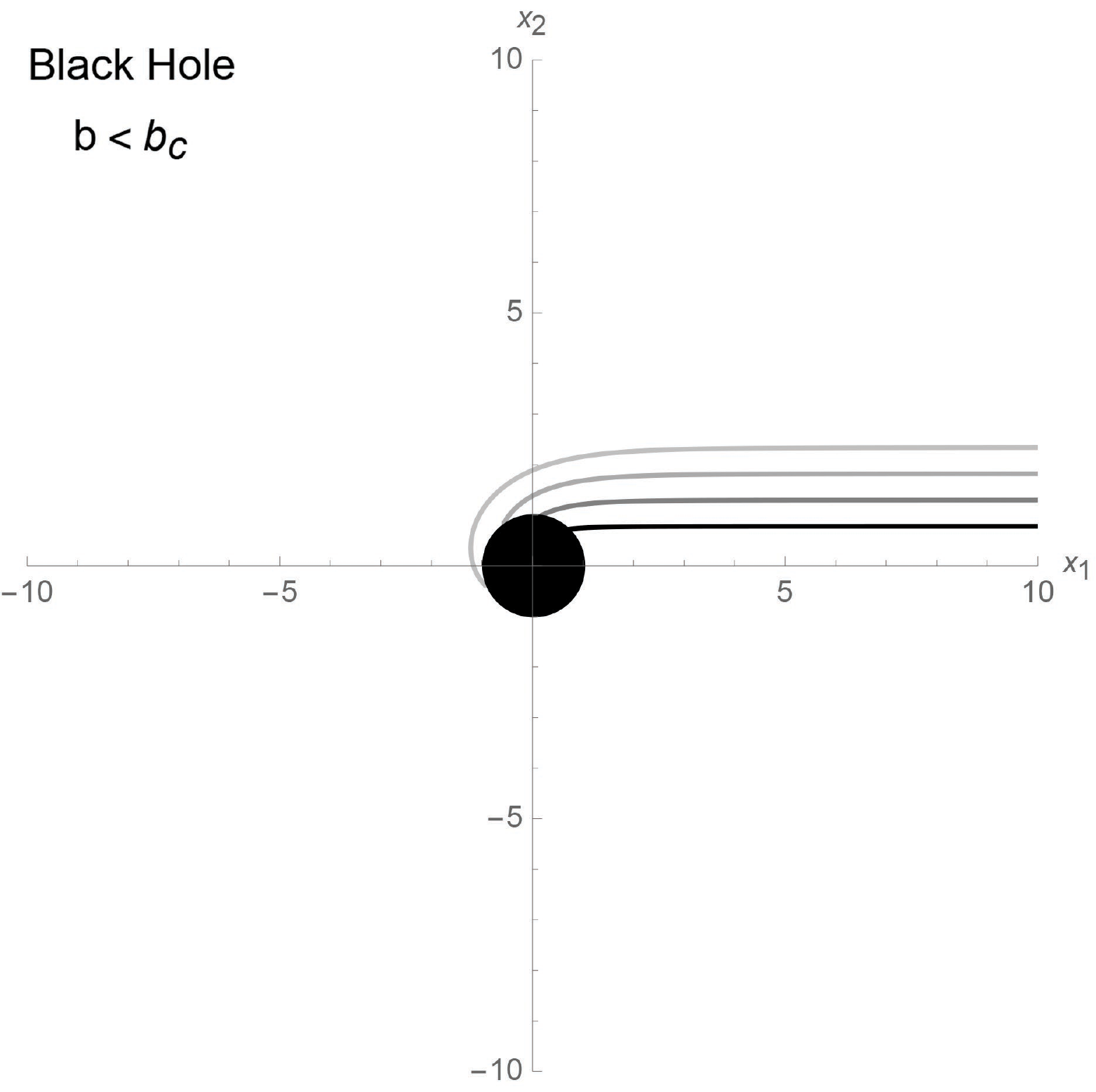} } \\
 {\includegraphics[scale=0.28]{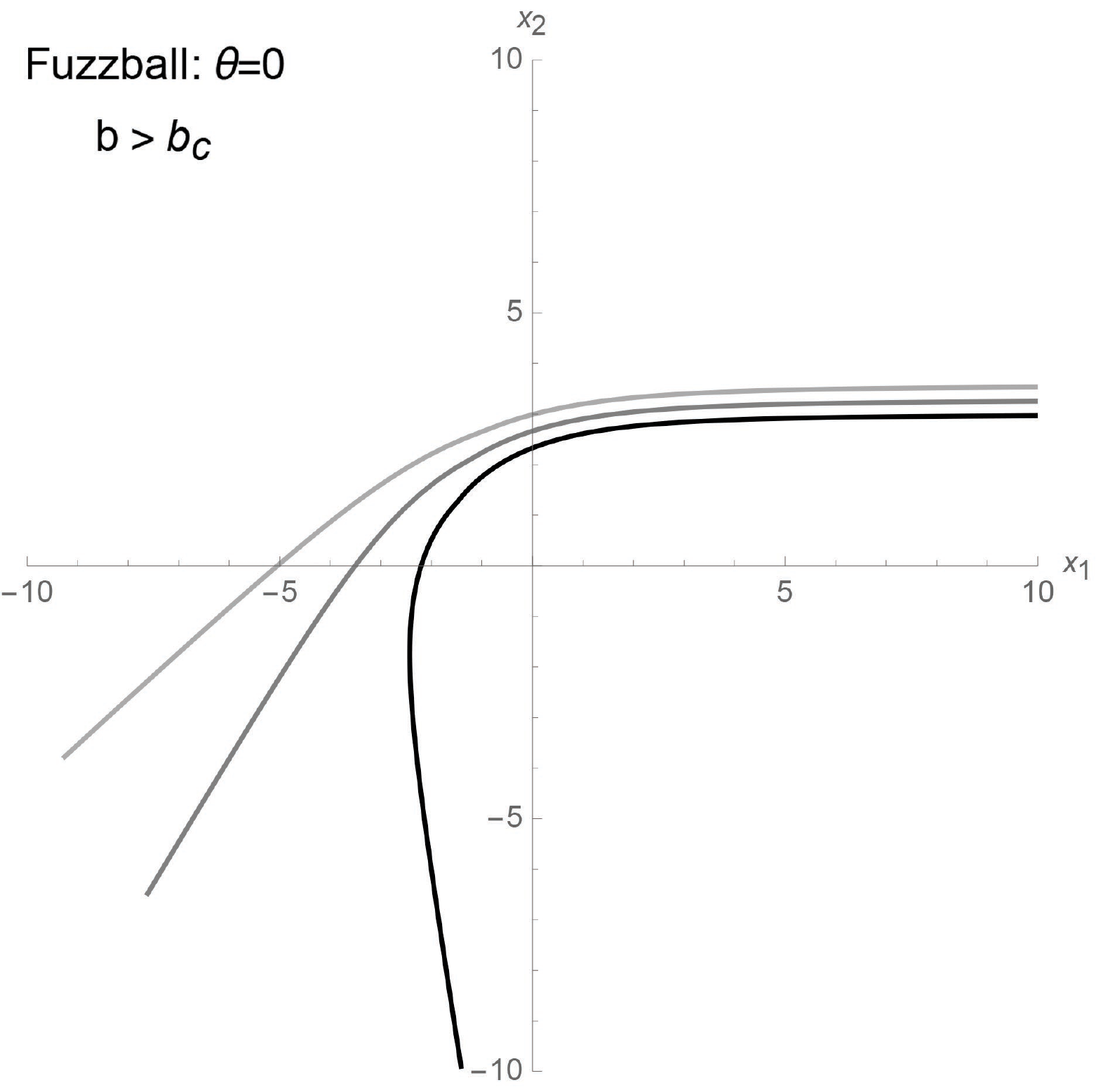} }
\quad
{\includegraphics[scale=0.28]{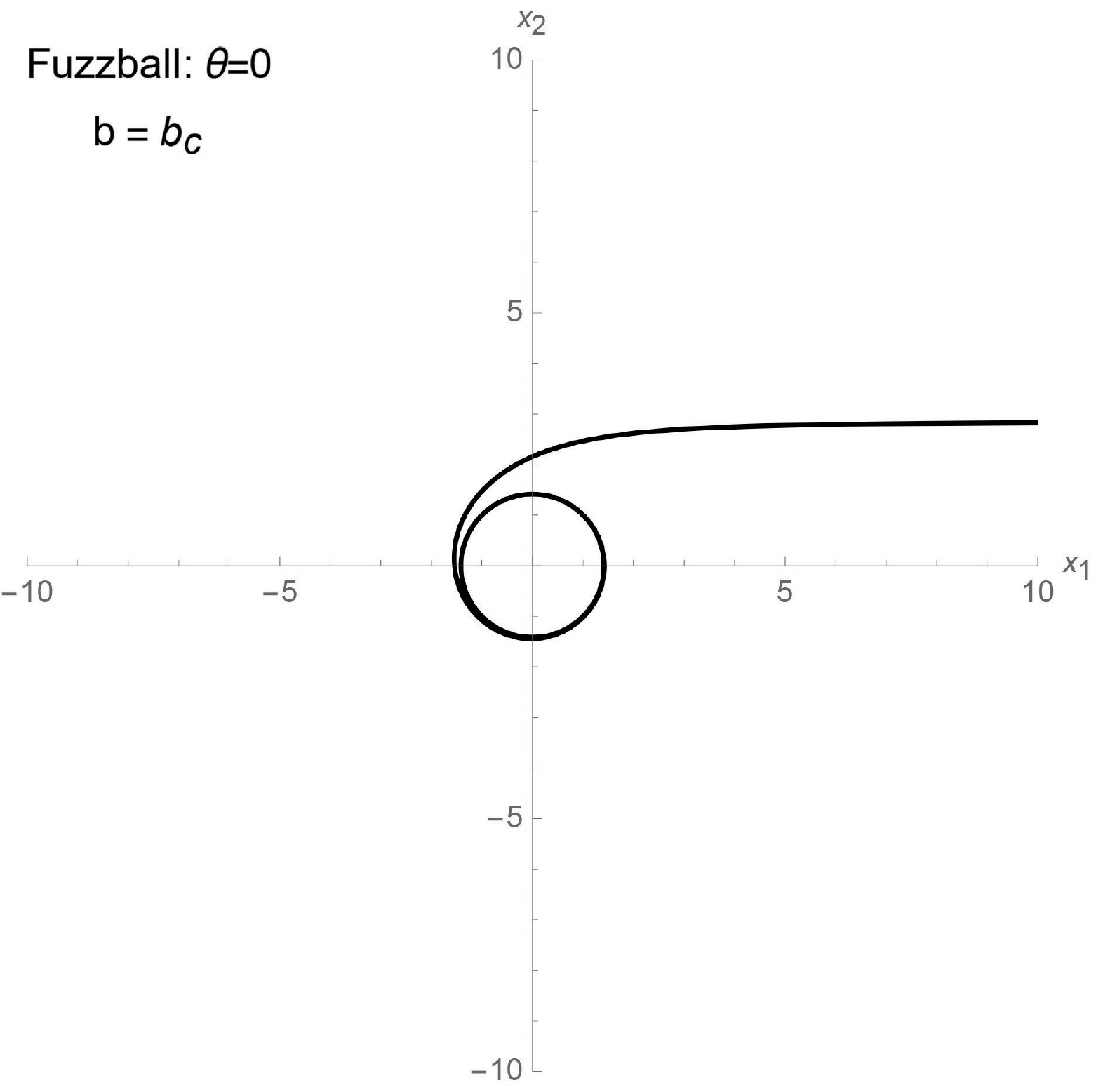} }
\quad
{\includegraphics[scale=0.28]{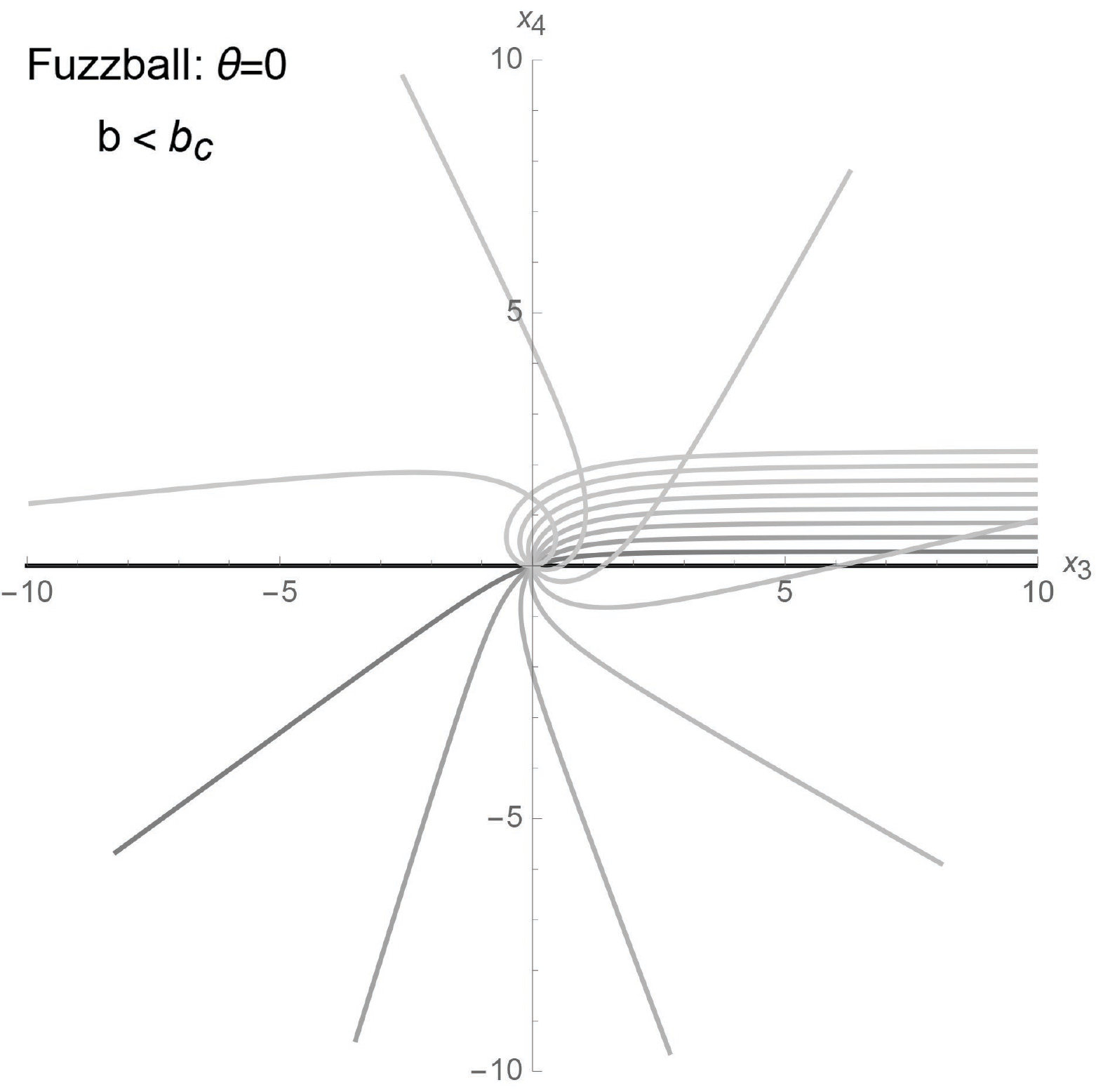} } \\
 {\includegraphics[scale=0.28]{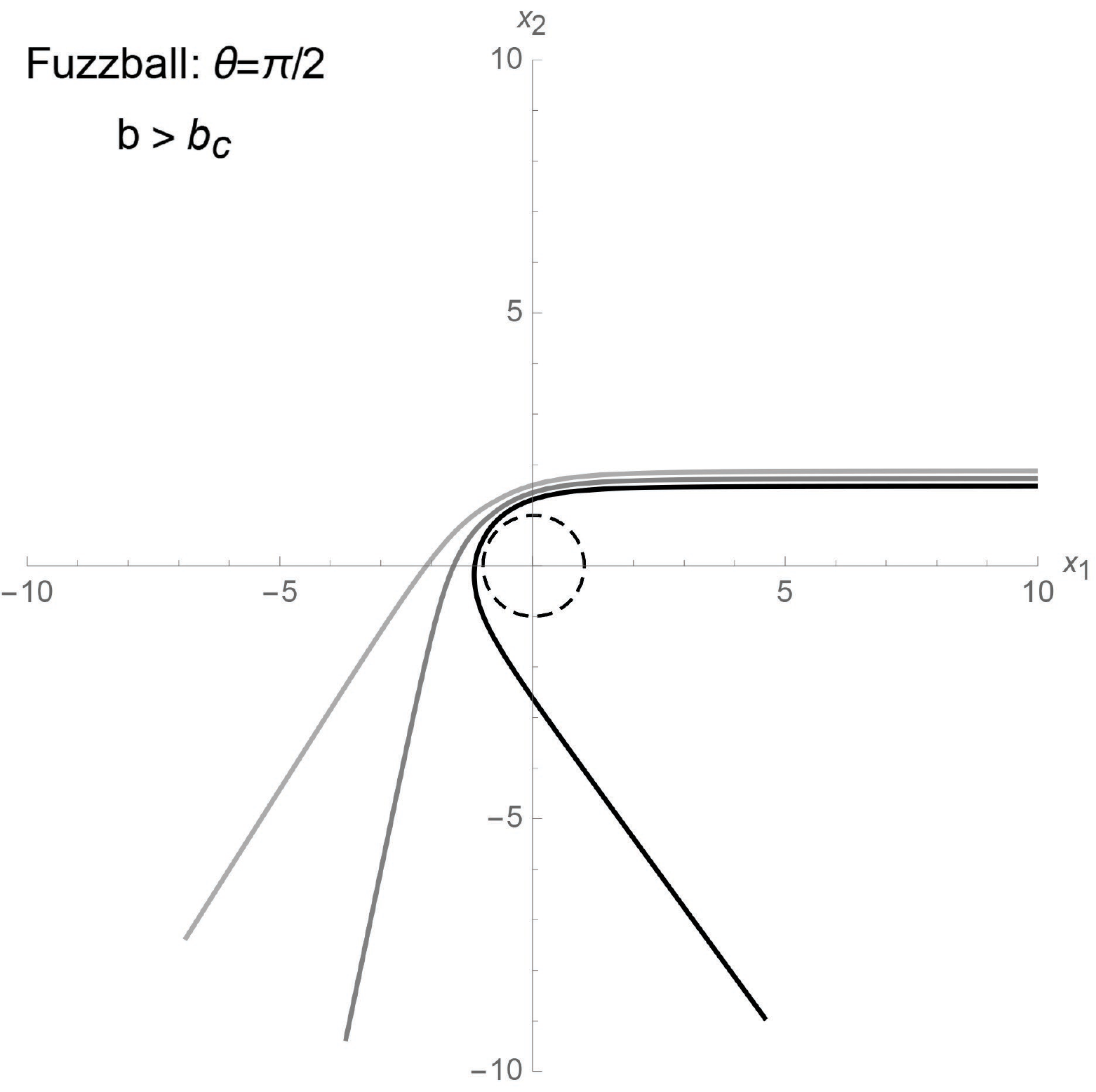}} 
\quad
{\includegraphics[scale=0.28]{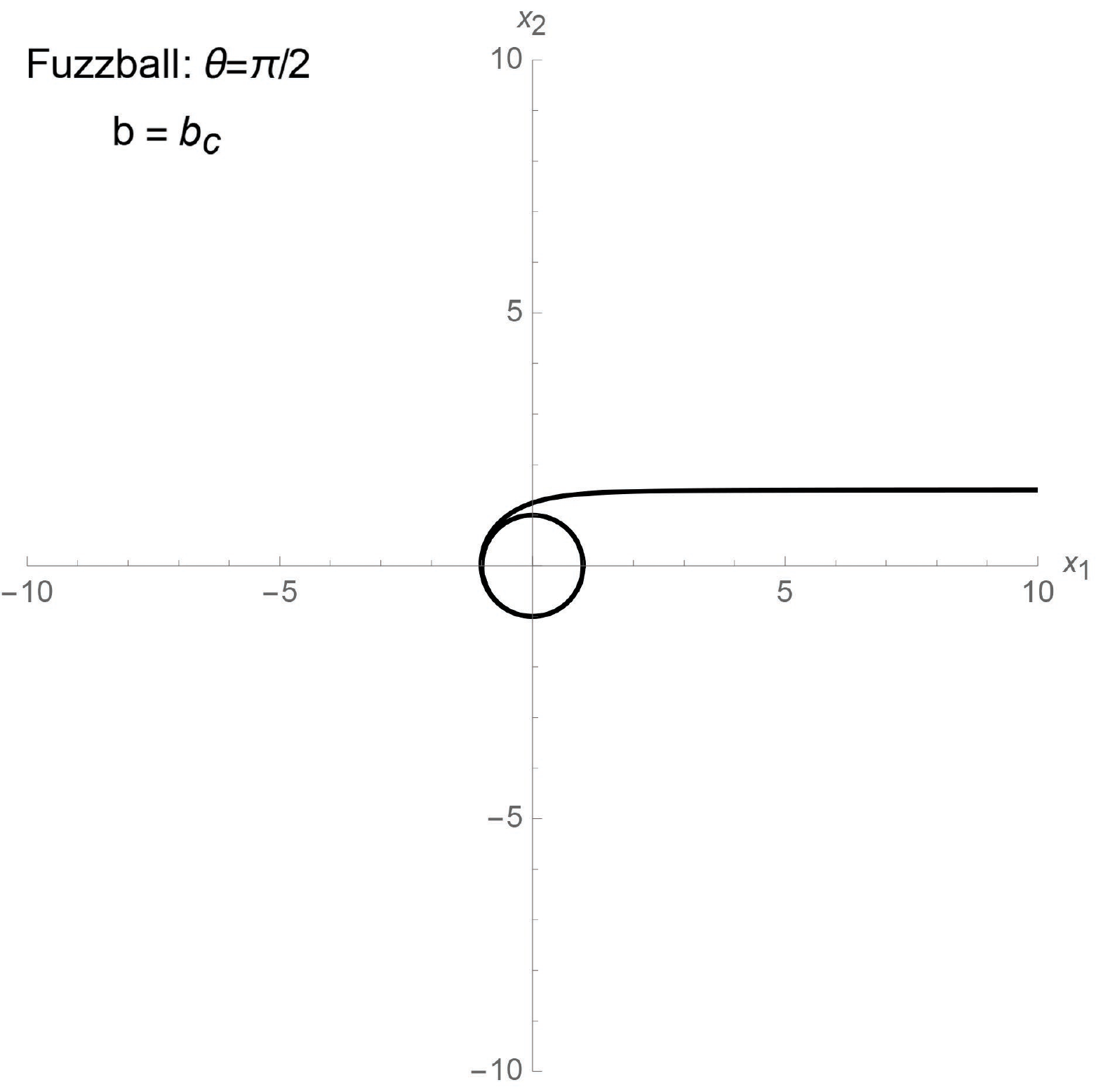} }
\quad
{\includegraphics[scale=0.28]{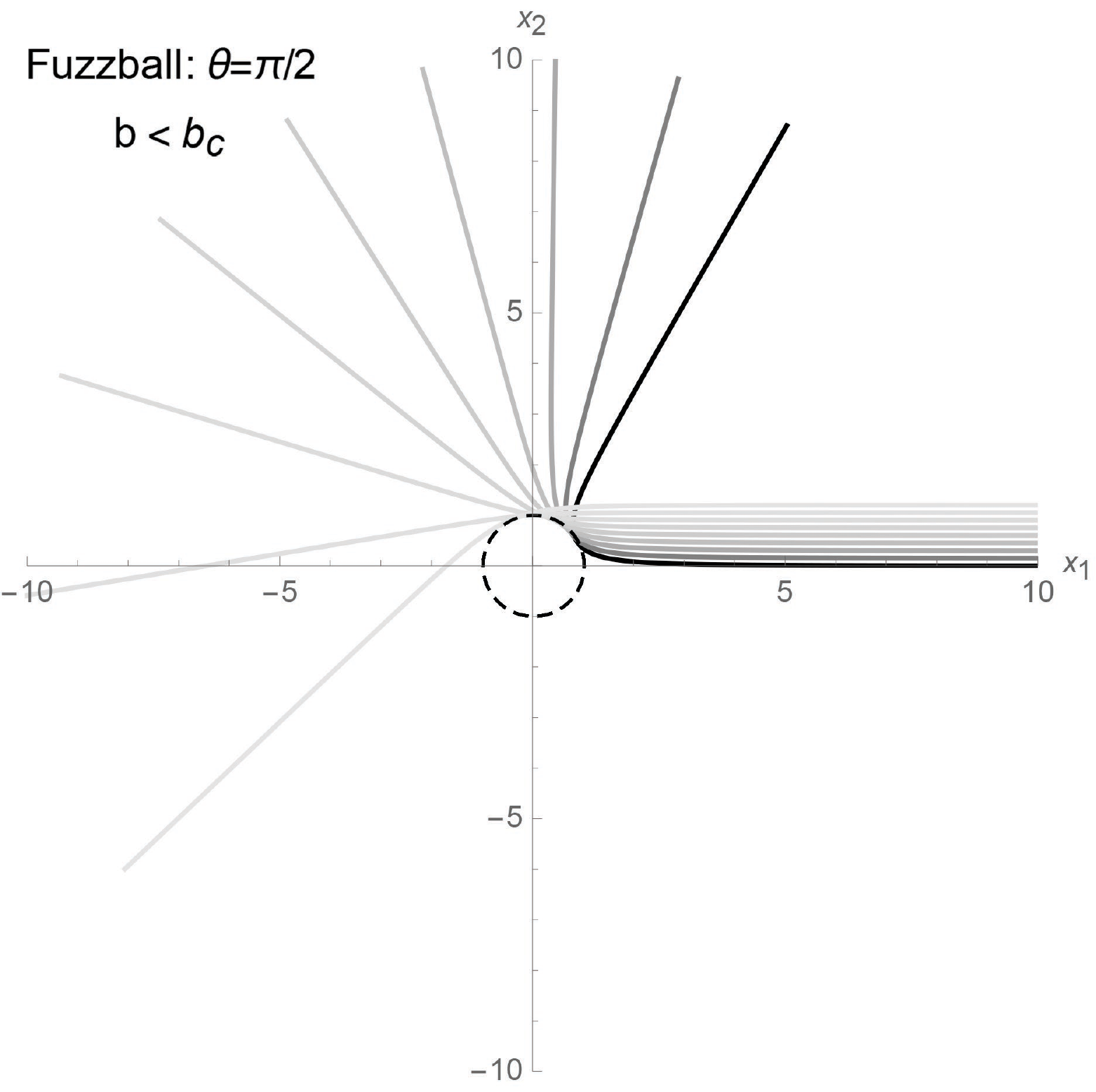} } 
\caption{Geodesics in the black hole and fuzzball geometries for different values of the impact parameter $b$.  }
\label{plots}
\end{figure}
\end{center} 

When the system is integrable and all the variables can be explicitly expressed in terms of $\rho$, the time (measured by an observer at infinity) required by a geodesic to reach the inversion (or turning) point $\rho_*$ starting from a point $\rho_0$ is given by
\footnote{The derivative of the time coordinate $t$ w.r.t. the affine parameter $\tau$ is given by
\be 
\frac{dt}{d\tau} = - \frac{\left(1 {-} \gamma_v\right)P_u {+} \widehat{P}_v}{\sqrt{2}} - \frac{1}{\sqrt{2}Z^2}\left[\gamma_\vartheta \widehat{P}_\vartheta + (\rho^2 {+} a^2)\gamma_\rho \widehat{P}_\rho + \frac{\beta_\psi(1{-}\gamma_v) {+} \gamma_\psi}{\rho^2 \cos^2\vartheta}\widehat{P}_\psi + \frac{\beta_\varphi(1{-}\gamma_v) {-} \gamma_\varphi}{(\rho^2 {+} a^2)\sin^2\vartheta}\widehat{P}_\varphi\right]
\ee} 
\be
\Delta t = \int_{\rho_0}^{\rho_*}  d\rho\,\left(\frac{dt}{d\tau}\right)\frac{\rho^2+a^2c_{\vartheta}^2(\rho)}{\rho^2+a^2}\,\frac{Z^2(\rho)}{\widehat P_\rho(\rho) }\label{TimeInt}
\ee
This integral may or may not diverge. Focusing for simplicity on geodesics with zero internal momenta ($P_y = 0$) and denoting by $K$ the total angular momentum of the incoming particle the impact parameter is given by $b=K/E$.
We can distinguish three distinct scenari depending on the value of $b$ (see figure \ref{plots}):
%Otherwise this integral does not make much sense since the condition $\widehat{P}_\rho$ is a function of all the dynamical variables and at best identifies a co-dimension one space. 
%Let us consider a massless infalling particle with definite total angular momentum $K$, see (\ref{Kblackstring}) below, its impact parameter is $b=K/\sqrt{E^2-P_y^2}$.
%For simple integrable systems, after expressing  $x^m(\rho)$ as a function of $\rho$, 

\begin{itemize}

 \item{ Scattering processes: They occur where either the geodesics encounter a turning point  $\rho_*>0$, i.e. a single zero of  $\widehat P_\rho^2(\rho)$ or when $\widehat P_\rho(\rho)$ is positive 
 everywhere and the time to reach $\rho=0$ is finite. This includes all  geodesics on black hole geometries  with large enough impact parameter and generic geodesics in fuzzball geometries.}
 
\item{Critical falling: They occur when geodesics encounter a {\it critical point} $\rho_*$  defined as a double zero of  $\widehat P_\rho^2(\rho)$. In this case, the time to reach $\rho_*$ is infinite and the particle asymptotically approaches $\rho_*$ without ever reaching it. This class of geodesics exists for specific choices of the impact parameter, both for black holes and fuzzballs. }  

 \item{Absorption processes: They occur for black hole geometries when geodesics find no turning point before the black hole horizon. In this case $\widehat P_\rho(\rho)$ is positive everywhere and the time to reach the horizon is infinite.
}

\end{itemize}

%When the system is not integrable, one can try and identify a `nearby' integrable system, {\it e.g.} by setting one charge to be smaller than the other ones and treating the corresponding terms as a `small' perturbation.

\section{Black hole geometry}
\label{BHnofuzz}

In this section we consider massless geodesics in the 3-charge five-dimensional  black hole geometry with and without angular momenta.

\subsection{The non-rotating three charge black hole} 
 
The non-rotating 3-charge black hole metric is obtained by taking  $a=n=0$ in (\ref{ds10}) and (\ref{metricQ1Q5Qp}). The $Z$-functions and one-forms  reduce to
\bea
Z_1 & =&1+\frac{L^2_1}{\rho^2  }  \qquad ,\qquad 
Z_2 = 1+\frac{L^2_5}{\rho^2 }  \qquad ,\qquad     Z^2=Z_1 Z_2   \nn\\
         \gamma_m dx^m &=& {\cal F}_0\, dv= -\frac{ L_p^2}{\rho^2} \,dv   \qquad ,  \qquad      \beta_m  = 0 
\eea
For this choice the oblate radius $\rho$ coincides with the spherical radius $r$ everywhere and the solution is spherically symmetric. The solution corresponds to a non-rotating five-dimensional black hole with a horizon at $\rho=0$ \cite{Gibbons:1982ih, Gibbons:1987ps}

 The `dressed' D1-brane charge $Q_1$, D5-brane charge $Q_5$ and Kaluza-Klein momentum $Q_P$ are given by
\be
\label{2Qangularmom}
Q_1=L_1^2 \quad, \quad Q_5=L_5^2 \quad, \quad Q_P=L_p^2 \, .
\ee
 The massless geodesic equation $\mathcal{H} =0$  can be written in the separable form
\be
2 \rho^2 Z^2 \mathcal{H} =  \left[  -  2 \rho^2 Z^2  \, P_u\, (P_v - {\cal F}_0 P_u)   +  \rho^2  P_\rho^2   \right] 
+  \left[  P_{\vartheta}^2   +\frac{  P_\varphi^2 }{  s_{\vartheta}^2}
+\frac{  P_\psi^2  }{  c_{\vartheta}^2}  \right]  =0    \label{h0bh}
\ee
where the two brackets account for $\rho$ and $\vartheta$ dependent terms, respectively.
The former equation can be solved by imposing that the combinations inside the brackets be constant, i.e.
\be 
K^2 =  P_{\vartheta}^2  +\frac{P_\varphi^2}{s_{\vartheta}^2}+ \frac{P_\psi^2}{c_{\vartheta}^2} =    2 \rho^2 Z^2  \, P_u\, (P_v - {\cal F}_0 P_u)   -  \rho^2  P_\rho^2  
\label{Kblackstring}
\ee
The right hand side equation can be solved for $P_\rho$
\be 
P_\rho^2 = -\frac{K^2}{\rho^2} + \frac{2P_u(\rho^2 + L_1^2)(\rho^2 + L_5^2)}{\rho^4} \left(P_v +  {L_p^2  P_u \over \rho^2}  \right)   \label{prhobh}
\ee
We notice that for 
\be
K^2 <  2  P_u^2   L_p^2  + 2 P_u P_v (L_5^2+L_1^2)   
\ee
the function $P_\rho^2$ is positive everywhere, so the geodesics extend down to the horizon at $\rho=0$.   The flight time down to the horizon diverges
\be
\Delta t \approx -L_1 L_5 L_p \int_{\rho_0}^{0}  {d\rho\over \rho^3} \,     \label{intdtBH}
\ee
as expected for a  black hole geometry.  
% For later comparison we notice that, according to (\ref{prhobh}),  the energy of the system can be written in the form 
%\be 
%E^2= {\rho^6 \over  (\rho^2 + L_1^2)(\rho^2 + L_5^2) (\rho^2 +  L_p^2 )  } \left[  P_\rho^2 +\frac{K^2}{\rho^2}  \right]
%\ee

\subsection{The rotating supersymmetric black hole}

   The analysis of geodesics in more general black hole backgrounds, extremal or not,  with or without charges and angular momenta, follows {\it mutatis mutandis} the same steps as before and the existence of a critical value for the total angular momentum of the incoming particles can be always displayed. In this section, we illustrate this universal feature by considering scattering from a three equal charge supersymmetric black hole with non-trivial angular momentum in five dimensions.   
   The metric of this black hole reads \cite{Breckenridge:1996is}
     \begin{equation}
\label{fvdmnsnlextrmblckhl}
\begin{aligned}
ds^2_S &= -\left(1 - \frac{\mu}{r^2}\right)^2\left(dt - \frac{\mu\omega\sin^2\vartheta}{r^2 - \mu}d\varphi -\frac{\mu\omega\cos^2\vartheta}{r^2 - \mu}d\psi\right)^2
+
\\
&+
\left(1 - \frac{\mu}{r^2}\right)^{-2}dr^2 + r^2\left(d\vartheta^2 + \sin^2\vartheta\,d\varphi^2 + \cos^2\vartheta\,d\psi^2\right)
\end{aligned}
\end{equation}
where $\mu$ is the mass parameter and $\omega$ accounts for the angular velocity. For concreteness, we focus on geodesics at constant $\vartheta$, let us say $\vartheta=0$\footnote{The analysis for $\vartheta=\pi/2$ is identical exchanging $\varphi \leftrightarrow \psi$}.  Consistently, we set
 $\dot{\vartheta}=\dot{\varphi}=0$, i.e. $P_\vartheta=P_\varphi=0$. 
   The corresponding Hamiltonian reduces to
\begin{equation}
\mathcal{H} = \ft12 g^{mn} P_m P_n=-\frac{1}{2}\left(1 - \frac{\mu}{r^2}\right)^{-2}E^2+ \frac{1}{2}\left(1 - \frac{\mu}{r^2}\right)^{2}P^2 + \frac{1}{2r^2}\left(J - \frac{\mu\omega E}{r^2 - \mu}\right)^2
\end{equation}
with
\begin{equation}
\begin{aligned}
-E &= g_{tn} \dot{x}^n = -\left(1 - \frac{\mu}{r^2}\right)^2\left(\dot{t} - \frac{\mu\omega}{r^2 - \mu}\dot{\psi}\right)
\\
J &= g_{\psi n}  \dot{x}^n= \frac{\mu\omega}{r^2 - \mu}\left(\dot{t} - \frac{\mu\omega}{r^2 - \mu}\dot{\psi}\right) + r^2\dot{\psi}\\
 P &=g_{r n}  \dot{x}^n= \left(1 - \frac{\mu}{r^2}\right)^{-2}\dot{r}\,.
 \end{aligned}
\end{equation}
The momenta $E$ and $J$ are conserved while $P$ is determined by solving  the null condition 
$\mathcal{H}=0$ leading to (in the incoming branch)
\begin{equation}
P(r^2) = - { r \over  \left(r^2 - \mu\right)^{2}} \left[E^2\, r^6 -  \left[J(r^2 - \mu) - \mu\omega E\right]^2\right]^{\frac{1}{2}}
\end{equation}
%Setting $\dot{t}=1$ the geodesic equation becomes 
%\begin{equation}
%{dr\over d\psi} = { r^2 P(r^2) \over \left(1 - \frac{\mu}{r^2}\right) \left(J - \frac{\mu\omega}{r^2 - \mu}\right)}
%\end{equation}
   We notice that, if $\omega^2<\mu $, the polynomial inside the brackets is positive for large $r$ and negative for $r=\sqrt{\mu}$ and therefore it vanishes for some $r_*>\sqrt{\mu}$. For this choice, the particle either bounces back or gets trapped inside a critical trajectory before it reaches the horizon at $r=\sqrt{\mu}$.  The trapping behaviour occurs if $J=J_c$ such that a point $r_*$ exists where $P(r_*)=P'(r_*)=0$.  Parametrising the angular momentum by means of the impact parameter $b=J/E$, the two equations are solved by taking 
 \be
 r_*= \Big| {2\,b_c\over 3 }\Big|
 \ee
with $b_c$ a solution of the cubic equation
\begin{equation}
4 b_c^3 - 27  \mu(b_c +\omega ) = 0
\end{equation}
The solutions are
\begin{equation}
b_c = 
- 3\sqrt{\mu}\,\sin\left(\frac{1}{3}\arctan\frac{\omega}{\sqrt{\mu - \omega^2}} + \frac{2\pi}{3}\,C\right) \label{jtree}
\qquad
,
\qquad
C = -1,0,1
\end{equation}
 It is easy to see that $C = 0$ leads to a zero $r_*<\sqrt{\mu}$  inside the horizon, so it should be discarded. The remaining two roots lead to critical geodesics of the black hole geometry. 
\begin{center}
\begin{figure}[t]
\centering
\includegraphics[scale=0.47]{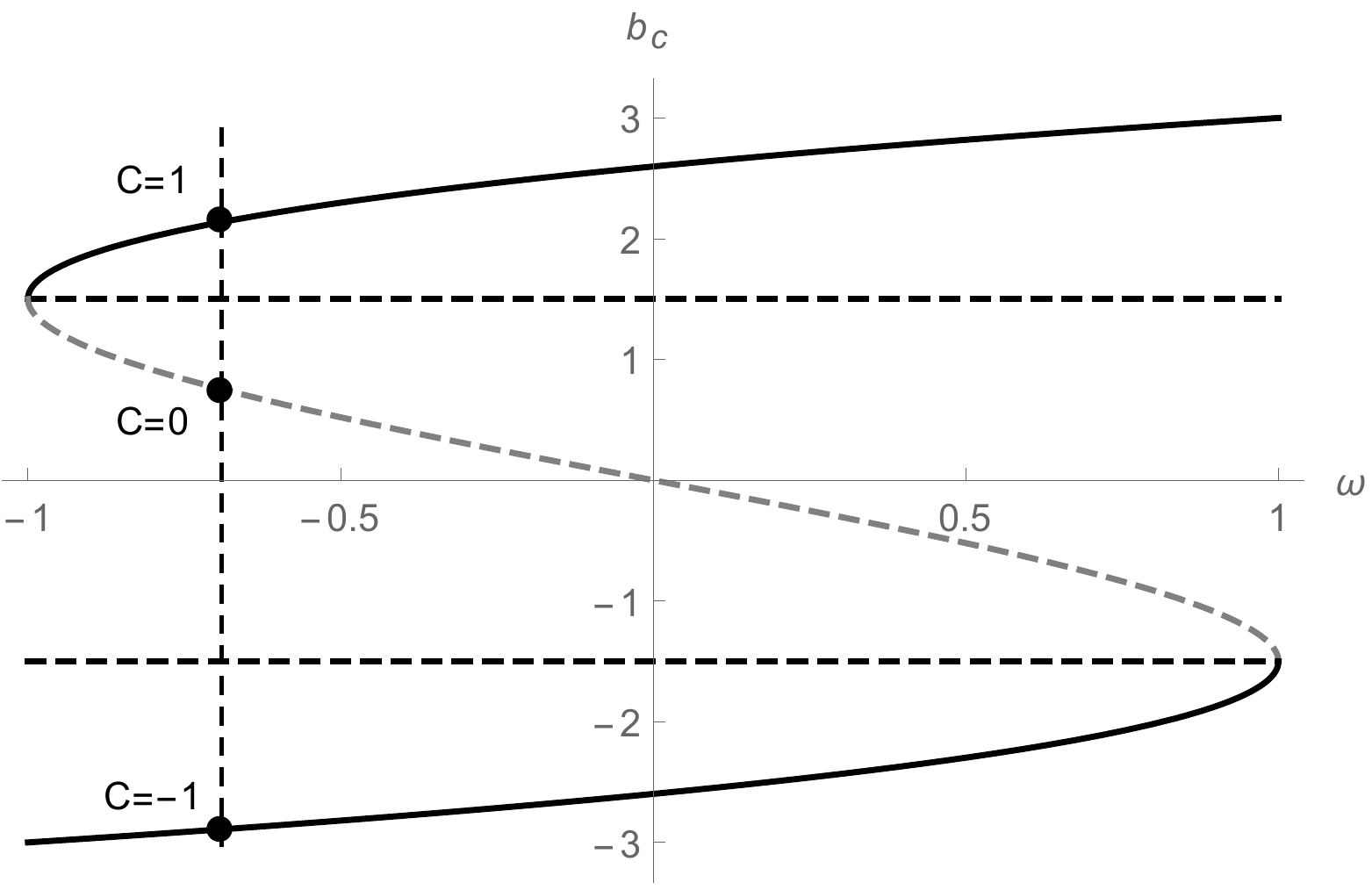}
\caption{Critical impact parameter $b_c$ vs the BH angular velocity $\omega$, both in units of $\sqrt{\mu}$. For every value of $\omega$ we find two different critical parameters, corresponding to the intersections with the solid line.}
\end{figure}
\end{center}  
\section{Two-charge fuzzballs}
\label{2chargefuzz}

 In this section we consider massless geodesics along 2-charge fuzzball geometries obtained by setting $\varepsilon_1=\varepsilon_4=n=0$ in the three-charge fuzzball solution.

\subsection{The circular fuzzball solution}   
  
  The general 2-charge geometry is specified by a profile function   $\vec{F}(v)$ with values on $\mathbb{R}^4\times \mathcal{T}^4$. 
Here we choose a circular profile $\vec{F}(v)$ in $\mathbb{R}^4$
\be
\vec{F}(v) = a\left(\cos\frac{2\pi v}{\lambda}\,,\sin\frac{2\pi v}{\lambda}\,,0\,,0\right)
\ee
for which one has
\be
\begin{aligned}
Z_1 &= 1 + \frac{L_1^2}{\lambda}\int_0^\lambda \frac{  \left|\dot{\vec{F}}(v)\right|^2\, dv}{\left|\vec{{X}}-\dot{\vec{F}}(v)\right|^2}=1+\frac{L^2_1}{\rho^2+a^2c_{\vartheta}^2 }
\\
Z_2 &= 1 + \frac{L_5^2}{\lambda}\int_0^\lambda \frac{dv}{\left|\vec{{X}}-\dot{\vec{F}}(v)\right|^2}=1+\frac{L^2_5}{\rho^2+a^2c_{\vartheta}^2} 
\end{aligned}
\ee
and $Z^2 =Z_1 Z_2$. Moreover the 1-forms $\beta$ and $\gamma$ are given by \cite{Lunin:2001fv}
 \bea 
 \beta  &=&\beta_m dx^m = \frac{ a^2 \,R }{  \rho^2+a^2c_{\vartheta}^2}\left(  \,s_{\vartheta}^2d\varphi -  c_{\vartheta}^2d\psi   \right)     \: , \:  \nn\\
 \gamma   &=& \gamma_m dx^m =\frac{ a^2 \, R}{  \rho^2+a^2c_{\vartheta}^2}\left( s_{\vartheta}^2d\varphi +  c_{\vartheta}^2d\psi   \right) 
\eea
% and 
% \be 
% \vec{A}(\vec{{X}})=\frac{L_1^2}{\lambda}\int_0^\lambda dv\,\frac{\dot{\vec{F}}(v)}{\left|\vec{{X}}-\dot{\vec{F}}(v)\right|^2}
%\ee
with  $R=  R_y / \sqrt{2} $,  $R_y$ being the radius of $S^1$ along the $y$-direction. 
The geometry has no horizon for
\be
a^2 =  \frac{L^2_1L^2_5}{2 R^2} 
\ee

\subsection{The geodesic equations}

The Hamiltonian depends only on ${\vartheta}$ and $\rho$, so the momenta $P_u$, $P_v$, $P_\psi$ and $P_\varphi$ are all conserved. 
The Hamiltonian can be separated \cite{Cvetic:1996xz,Cvetic:1997uw,Lunin:2001dt,Chervonyi:2013eja} according to   
\bea
 2 Z^2  \, (\rho^2+a^2 c_{\vartheta}^2) \,\mathcal{H} =\lambda_\rho (\rho,P_\rho) +\lambda_{\vartheta} ({\vartheta},P_{\vartheta})   \label{nulllambda}
\eea
with
\bea
\label{fuzzgeodsepKlambda}
\lambda_{{\vartheta}} ({\vartheta},P_{\vartheta})   &=&P_{{\vartheta}}^2 + {P_\psi^2 \over \cos^2{\vartheta}} + {P_{\vartheta}^2 \over \sin^2 {\vartheta}} + 2 a^2 \sin^2 {\vartheta} \, P_u P_v \\
\lambda_{\rho} (\rho,P_\rho) &=&
 ({\rho^2{+}a^2})P_{\rho}^2+{a^2  \widetilde{P}_\psi^2  \over\rho^2}- { a^2 \widetilde{P}_\varphi^2 \over\rho^2 + a^2}  - 
2 ({\rho}^2{+}a^2{+}L_1^2{+}L_5^2) P_u P_v  
\eea
and
\be
\widetilde{P}_\psi = P_\psi  +R\, (P_v-P_u)  \qquad , \qquad
\widetilde{P}_\varphi = P_\varphi   + R\, (P_v+P_u) 
\ee

Equation $\mathcal{H} =0 $ can be solved by taking 
\be
\lambda_{{\vartheta}}=-  \lambda_{\rho} = K^2 
\ee
with $K$ a constant, that can be interpreted as the total angular momentum. Equivalently one has
 \bea
\label{fuzzgeodsepK}
P_{{\vartheta}}({\vartheta})^2  &=& K^2- {P_\psi^2 \over c_{\vartheta}^2 } - {P_\varphi^2 \over s^2_{\vartheta}} -  2 P_u P_v \,a^2  s_{\vartheta}^2     \nn \\
P_{\rho}(\rho)^2 &=&  -{a^2 \widetilde{P}_\psi^2  \over\rho^2({\rho^2{+}a^2})  }+ { a^2 \widetilde{P}_\varphi^2 \over (\rho^2 + a^2)^2}  +
{  2 \left({\rho}^2{+}L_1^2{+}L_5^2+a^2 \right) P_u P_v     -K^2    \over  {\rho^2{+}a^2}  }    
\eea
Expressing the velocities in terms of the momenta 
\be
\label{eomfortheta}
\nn
\dot{{\vartheta}} ={ P_{\vartheta}({\vartheta}) \over Z^2 (\rho^2+a^2 c_{\vartheta}^2) }\quad ,
\qquad
\dot{\rho} = { \rho^2+a^2\over \rho^2+a^2 c_{\vartheta}^2 }
{ P_\rho(\rho)\over Z^2 }
\ee
one finds the separable geodesic equation
\bea
{d{\vartheta}\over P_{\vartheta} ({\vartheta}) } ={ d\rho\over P_\rho (\rho) (\rho^2+a^2) } 
\eea
  that implicitly determines ${\vartheta}(\rho)$ in terms of elliptic integrals. 
  Finally, $\varphi(\rho)$ and $\psi(\rho)$ follow from
 \be
d\psi  = { \rho^2 P_\psi+a^2 c^2_{\vartheta}\, \widetilde{P}_\psi  \over P_\rho(\rho)  \rho^2  (\rho^2+a^2)  c_{\vartheta}^2}   \, d\rho\quad ,
\quad 
d\varphi  = { (\rho^2+a^2)P_\varphi -a^2\, s^2_{\vartheta}   \widetilde{P}_\varphi   \over P_\rho(\rho)  \rho^2  (\rho^2+a^2) s_{\vartheta}^2 }   \, d\rho
\ee
after integration over $\rho$.

%
%\begin{center}
%\begin{figure}[t]
%\centering
%\subfloat[][2-plane $\vartheta = \pi/2$, the dashed circle is the circular profile.]{
%%\includegraphics[scale=0.3]{2-charge_plane_12_b_leq_b_c}\label{Fig2q_a}}
%\includegraphics[scale=0.3]{fig_Fuzz1bless}\label{Fig2q_a}}
%\qquad
%\subfloat[][2-plane $\vartheta = 0$.]{
%%\includegraphics[scale=0.3]{2-charge_plane_34_b_leq_b_c}\label{Fig2q_b}}
%\includegraphics[scale=0.3]{fig_Fuzz0bless}\label{Fig2q_b}}
%%\caption{Scattered geodesics in the 2-charge fuzzball geometry for $b < b_c$.}
%\label{Fig2q}
%\end{figure}
%\end{center}
%

\subsection{Critical geodesics}

 It is convenient to write
 \be
 P_\rho^2 (\rho)=  {{\cal P}_3(\rho^2) \over  \rho^2 (   {\rho^2{+}a^2})^2  }  \label{p3c}
\ee
and set $\rho^2 = x$ so that
\be
\label{twochargespoly}
{\cal P}_3(x)=A \,x^3+B\, x^2+C \,x+D
\ee
with
\bea
A &=& 2\, P_u P_v \nn\\
B &= &    2P_u P_v(2a^2 + L_1^2+ L_5^2)      -K^2   \nn\\
C &=&  a^2 \left[\widetilde{P}_\varphi^2 - \widetilde{P}_\psi^2   + 2 P_u P_v (a^2 {+} L_1^2 {+} L_5^2)    -K^2  \right] \nn\\
D&=&   -a^4  \widetilde{P}_\psi^2   \label{abcd}
 \eea
Since $A>0$ and $D<0$, the  polynomial ${\cal P}_3(x)$ is positive for large $x$ and negative for small $x$. Therefore it has at least a zero $x_*$ 
(the largest one) for positive $x=\rho^2$. This is in contrast with the behaviour observed for the black hole geometry, where  $P_\rho^2 (\rho)$ was shown to be positive everywhere for small enough angular momenta $K$. 
We conclude that massless probes in the fuzzball metric escape from the gravitational background, even for low values of the angular momentum $K$. An exception occurs when the angular momentum is tuned such that $x_*$  is  a double zero of ${\cal P}_3(x)$ , {\it i.e.}  
   \be
   {\cal P}_3(x_*)=  {\cal P}_3'(x_*)=0     \label{xd}
   \ee
For this choice, the integral (\ref{TimeInt}) diverges and the surface  $\rho_* =\sqrt x_*$  looks like a horizon for the massless geodesics. Indeed, for a critical value of  $K$  such that the two largest roots of  ${\cal P}_3(x)$ collide, the particle winds around the target forever, asymptotically  approaching the `circular' orbit with radius $\rho_*$. Such geodesics will be referred to as {\it critical geodesics}. In the remaining of this section we will display some explicit choices of kinematics exhibiting such trapping behaviour.
 
 First, we notice  that the conditions  $A>0$ and $D<0$, together with the requirement that the largest root is double and positive,  imply that 
    all three roots are positive and
    \be
    A,C >0     \qquad , \qquad  B,D <0    \label{abcdin}
    \ee
Solving  (\ref{xd}) for $x_*$ and $D$ one finds
\be
\begin{aligned}
x_* &= \frac{1}{3 A} \left(-B + \sqrt{B^2-3 A C} \right)
\\
%\quad
%x_S = \frac{1}{3 A} \left(-B - 2\sqrt{B^2-3 A C} \right)
%\\
D &=\frac{2}{27A^2} (B^2-3 AC)^{3/2} - \frac{B}{27A^2}(2 B^2-9AC)
\end{aligned}
\ee
 Solutions compatible with (\ref{abcdin}) exist if 
 \be
 4 A C \geq B^2 \geq 3AC
 \ee
  The two extreme cases where the inequalities are saturated are easy to solve in analytic form:

\begin{itemize}
\item {Case I: $B^2=3AC$. For this choice all three roots collide and $D = \frac{BC}{9A}$. 
 From (\ref{abcd}) one finds 
\be 
\begin{aligned}
\widetilde{P}_\varphi^2 &= \frac{\left[ K^2+2 (a^2 - L_1 ^2 - L_5^2) P_u P_v   \right]^3 }{  108 \,a^4\, P_u^2 \, P_v^2	}
\\
\widetilde{P}_\psi^2 &= \frac{\left[ K^2 -2 (2a^2 + L_1 ^2 + L_5^2) P_u P_v  \right]^3 }{  108 \,a^4\, P_u^2 \, P_v^2	}
\end{aligned}
\ee
and
\be 
\rho_*^2 = \frac{K^2}{6 P_uP_v  } -\ft13 (2a^2 + L_1 ^2 + L_5^2)  >0 
\ee
  We notice that a critical geodesic of this type exists for a large enough total angular momentum  $K$. 
}

\item {Case II:  $B^2=4AC$. For this choice one finds $D=0$, 
\be 
\begin{aligned}\label{caseii}
\widetilde{P}_\psi &= 0
\\
~~~  \widetilde{P}_\varphi^2 &=  \frac{ \left[K^2-2 P_uP_v (L_1 ^2 + L_5^2) \right]^2}{8 a^2 P_u P_v}
\end{aligned}
\ee
and
\be 
\rho_*^2 =\frac{K^2}{4 P_uP_v  } -\ft12 (2a^2 + L_1 ^2 + L_5^2)  >0 
\ee
}
\end{itemize}

\subsection{An example of critical geodesics}

To illustrate the trapping behaviour of fuzzballs, let us consider the critical geodesics along the plane $\vartheta = \pi/2$, for the choice
\be
L_1=L_5=a \qquad , \qquad P_u=P_v \qquad , \qquad P_\psi=0
\ee
For this choice the velocity $\dot{y}$ of the particle along the compact circle can be set to zero along the full trajectory. 
The critical geodesics fall into case II  above. Introducing the impact parameter 
\be
b={P_{\varphi}  \over E}={ P_{\varphi} \over \sqrt{2} P_u}
\ee
 and using 
(\ref{p3c}), (\ref{abcd}), (\ref{fuzzgeodsepK})  
 one finds
 \be
 {\cal P}_3 (\rho)=2 P_u^2 \rho^2 \left[\rho^4+(3 a^2{-}b^2)\rho^2 + (3a{-}2 b) a^3\right] 
\ee
with largest zero
 \be
 \rho_*^2={ b^2-3 a^2 +  \sqrt{ (b-a)^3 (b+3a) } \over 2  }
 \ee 
The turning point exists for $b \leq -3a$ or $b \geq 3a/2$; when $b=3a/2$ or $b=-3a$ a limit cycle exists at $\rho=0$ and $\rho=\sqrt{3}\, a$ respectively. For values of $b$ in-between $P_\rho^2$ has no zeroes, the probe reaches $\rho=0$ in a finite, possibly large, amount of time, surpasses it and gets scattered back at infinity.
The time to reach $\rho_*$ is given by
 \be
\Delta t = \int_{\rho_0}^{\rho_*}  d\rho \,{ \rho^4+3a^2 \rho^2+(3a{-}b)a^3 \over \rho^2+a^2} \, { \sqrt{2} P_u \, \rho \over \sqrt{{\cal P}_3(\rho^2)} }    \label{intdt}
\ee

\begin{center}
\begin{figure}[t]
\includegraphics[scale=0.7, center]{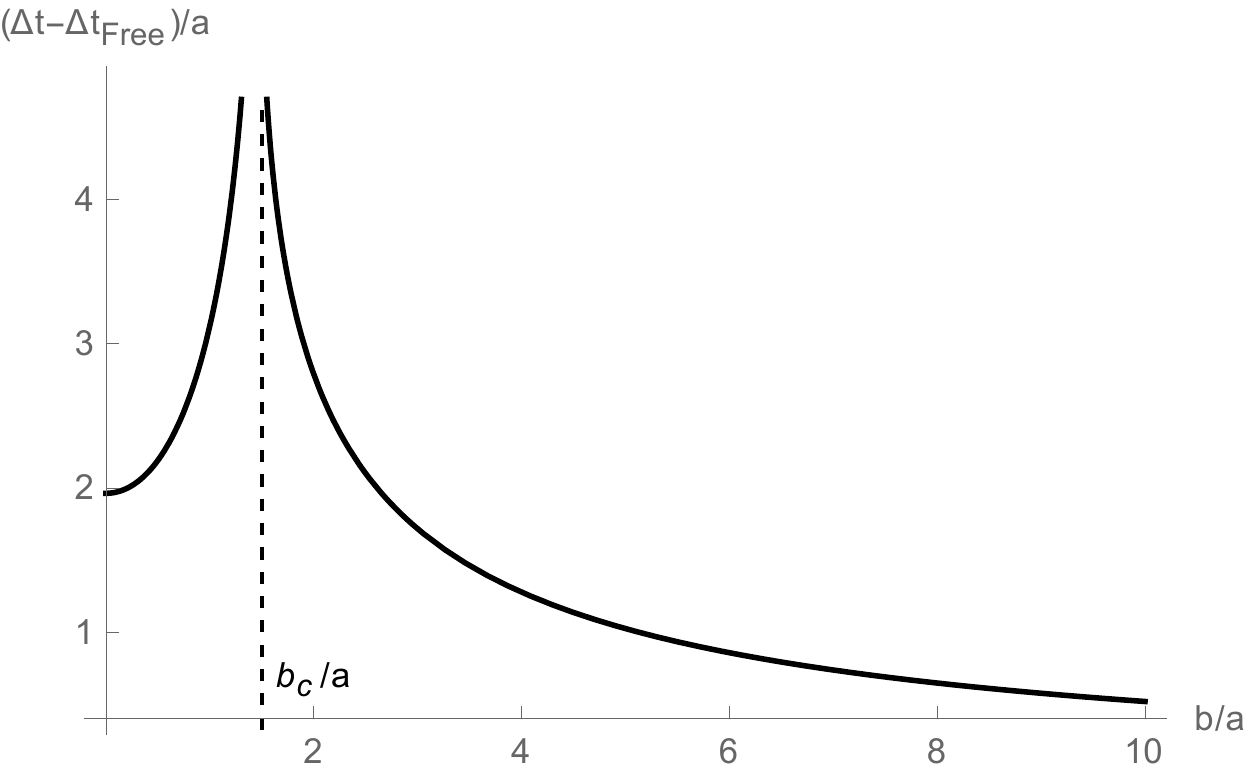}
\caption{Time delay between massless particles moving in a 2-charge fuzzball geometry and flat space-time as a function of the adimensionalised impact parameter $b/a$.}
\label{TimeDelay}
\end{figure}
\end{center}

In (Fig.$\,$\ref{TimeDelay}) we display the difference between the total flight time in the fuzzball geometry and in flat space-time as a function of $b$ for a fixed large $\rho_0$. As expected, the closer a particle's impact parameter approaches the critical one, the longer the time it will spend orbiting around the fuzzball. It is also clear that even though for $b < b_c$ the particle will eventually be scattered, it spends a considerable amount of time in the proximity of the fuzzball.
   
\section{3-charge fuzzballs}
\label{3chargefuzz}

In this section we consider scattering on 3-charge fuzzball  geometries.

\subsection{The  Hamiltonian and momenta  }

 Momenta and velocities in the 3-charge geometry are related  by
  \bea
P_u &=& -(\dot v +\beta_m \dot x^m) \qquad 
\widehat{P}_v = - (\dot u + \gamma_m \dot x^m )   \qquad 
\widehat{P}_\rho =  { Z^2(\rho^2+a^2c_{\vartheta}^2)\over \rho^2+a^2} \dot{\rho}   \nn\\
\widehat{P}_{\vartheta} &=&  Z^2(\rho^2+a^2c_{\vartheta}^2)\, \dot{{\vartheta}} \qquad 
\widehat{P}_\psi = Z^2 \rho^2 c_{\vartheta}^2 \, \dot \psi  \qquad 
\widehat{P}_{\varphi} = Z^2 (\rho^2+a^2) \,s_{\vartheta}^2   \dot\varphi    \label{momenta3Q}
\eea
   The important difference with respect to the 2-charge case is that now $\beta_m$, $\gamma_m$ and $Z$, and therefore the Hamiltonian, explicitly depend on the combination  $\phi=\varphi+{nv\over R}$ and therefore $P_v$ and $P_\varphi$ are no longer conserved separately but only their combination $P_\nu = P_v - {n\over R} P_\varphi$ is. Indeed, the equations of motion  become
       \bea
 \dot P_u &=& \dot P_\nu=\dot P_\psi ={\cal H}=0 \nn\\
 \dot P_{\vartheta} &=& -{\partial {\cal H} \over \partial {\vartheta}} \nn\\
   \dot P_\varphi &=& -{\partial {\cal H} \over \partial \varphi}  = -{R\over n} {\partial {\cal H} \over \partial v} ={R\over n} \dot P_v  
  \eea
    We observe that the Hamiltonian ${\cal H}$ is a rational function of $\cos{\vartheta}^2$ and therefore
    \be
    {\partial {\cal H} \over \partial {\vartheta}}\sim  \cos{\vartheta} \sin{\vartheta}
    \ee
    This implies that $P_{\vartheta}$ is conserved for ${\vartheta}=0,\pi/2$. Moreover at  ${\vartheta}=0,\pi/2$,   $\widehat{P}_{\vartheta}=P_{\vartheta}$ and therefore constant $P_{\vartheta}$ implies constant $\dot{{\vartheta}}$. 
 We conclude that geodesics starting at   ${\vartheta}=0,\pi/2$ with zero  initial ${\vartheta}$ velocity, $\dot{{\vartheta}} =0$ keep ${\vartheta}$ constant along the whole trajectory.  
      In the following we restrict ourselves on geodesics along these two planes.

\subsection{${\vartheta}=0$ geodesics}

Let us start by choosing $n=1$ and considering the geodesics in the plane ${\vartheta} = 0$, orthogonal to the circular profile. 
The functions and forms defining the metric assume the following expression
\bea
\label{eqnsthetazero}
Z_4 &=& 0 \nn\\
 \beta  & =& -\frac{ a^2\, R }{  \rho^2+a^2 }d\psi   \nn\\
 \gamma  &=&  \frac{ a^2\, R }{  \rho^2+a^2} \left( 1 - {\cal F }_1    \right) d\psi +  {\cal F}_1 dv \nn\\
   {\mathcal{F}}_1  &=&  -\frac{{\varepsilon}_4^2}{2 (\rho^2+a^2)}  \nn\\ 
 Z^2  &=&  Z_1 Z_2 = \left(1+{L_1^2\over \rho^2+a^2} \right)\left(1+{L_5^2\over \rho^2+a^2} \right)
\eea 
Taking $\widehat{P}_{\vartheta}=P_{\vartheta}=0$ and  $P_\varphi=\widehat{P}_{\varphi} = 0$, the Hamiltonian becomes
 \bea
  {\cal H}  &=&    -    \, P_u\, \widehat{P}_v   +{1\over 2Z^2} \left( \widehat P_\rho^2 
+\frac{\widehat  P_\psi^2  }{ \rho^2}   \right)   \label{hhplane34}
\eea 
with
 \bea
\widehat{P}_v &=& P_v+\frac{{\varepsilon}_4^2}{2 (\rho^2+a^2)}  P_u \qquad, \qquad  
\widehat{P}_\rho =  P_\rho   \nn\\
  \widehat{P}_\psi  &=&  P_\psi  -\frac{ a^2\, R }{  \rho^2+a^2 } (P_u-P_v)     \,. \nn
 \label{momenta3Qplane12}
\eea
Recall that $P_u$, $P_v$, $P_\psi$ are conserved quantities.  
Plugging this into (\ref{widerho})  one finds
  \be
   P_\rho  = \pm   \left[   2Z^2\,  P_u\, \widehat{P}_v   
-\frac{\widehat  P_\psi^2  }{ \rho^2  }  \right]^{1\over 2} = \pm { {\cal P}_4(\rho^2)^{1\over 2} \over  \rho (\rho^2+a^2)^{3\over 2}  } 
\ee
with, setting $\rho^2=x$ as above,
\bea
{\cal P}_4 (x)  &=&       P_u\, x\,  ( x{+}a^2{+}L_1^2)   ( x{+}a^2{+}L_5^2)   \left[ 2\, P_v (x{+}a^2) + {\varepsilon}_4^2 \,  P_u \right] \nn\\
&& \qquad - (x{+}a^2)\left[ P_\psi ( x{+}a^2 ) -a^2\, R (P_u{-}P_v)  \right]^2   
 \label{widerho2}
  \eea
  We notice that the polynomial ${\cal P}_4(x)$  is positive for $x \to \infty$ and negative for $x \to 0$. Therefore it has a zero somewhere on the positive $x$ axis. Again we denote $x_*$ the largest positive zero. If $x_*$ is simple then it is a turning point and the particle gets deflected in the gravitational background.  
  On the other hand for a critical choice of $P_\psi$ for which $x_*$ is a double zero the particle gets trapped in the gravitational background, asymptotically approaching $\rho_* =\sqrt{x_*}$.
   
As an illustration of this critical behavior, let us consider a  particle with no internal Kaluza-Klein momentum $P_v=P_u$ and
\be
L_1^2=L_5^2={\varepsilon}_4^2/2=L^2\geq 3 a^2\,.
\ee
For this choice the polynomial ${\cal P}_4(x)$ takes the simple form
\be
\begin{aligned}
\mathcal{P}_{4}(x)
= 2P_u^2\, x\, (x+a^2 +L^2)^3 - (x + a^2)^3 P_\psi^2  
\end{aligned}
\ee
 Solving the critical conditions $\mathcal{P}_{4}(x)=\mathcal{P}'_{4}(x)=0$ one finds a double zero at
 \be
 x_*=L^2-a^2+ L\sqrt{L^2-3 a^2}
 \ee
 for the critical choice of angular momentum
 \be
P_{\psi}=\sqrt{6}P_u L\left[ 1 + \frac{L^2}{9 a^2} - \frac{L^2}{9 a^2}\left(1-\frac{3a^2}{L^2}\right)^{3/2}\right] \label{bcrit}
 \ee
  In other words, scattering massless particles off the fuzzball geometry, one finds that the components  with $P_\psi$ satisfying (\ref{bcrit}) are missing in the out-going spectrum, and the fuzzball geometry behaves effectively as a black object for the selected ``channel". 
  
  \subsection{${\vartheta} = {\pi/2}$ geodesics}
  \label{3chpi2}
  
In this plane, the Hamiltonian, explicitly depends on the combination  $\phi=\varphi+{nv\over R}$, so it is convenient to introduce the canonically related variables $\phi$, $\nu$ (and their conjugate momenta)
 \bea
\varphi  &=&    \phi-  {n v\over R}    \qquad , \qquad  P_\varphi= P_\phi  \nn\\
v  &=&   \nu    ~~~~~~~~\qquad , \qquad    P_v = P_\nu + {n\over R}\, P_\phi
 \eea
    In terms of these variables the equations of motion  become
       \bea
 \dot P_u &=& \dot P_\nu=\dot P_\psi ={\cal H}=0 \nn\\
 \dot P_{\vartheta} &=& -{\partial {\cal H} \over \partial {\vartheta}} \nn\\
  \dot P_\phi &=& -{\partial {\cal H} \over \partial \phi} 
  \eea

 For motion  in the plane of the string profile, the metric is given by (\ref{ds10}) and (\ref{metricQ1Q5Qp}) with
\bea
Z_1 & =&1+\frac{L^2_1}{\rho^2  }+ \frac{\,{\varepsilon}_1R^2 \, \Delta_n \, \cos 2\phi  }{   L_5^2 \rho^2  } 
\qquad\qquad\qquad\,~~ \nn
Z_2 = 1+\frac{L^2_5}{\rho^2 }  
\\
Z_4^2 &=&{ 2\, {\varepsilon}_4^2 R^2  \Delta_n \, \cos^2 \phi \over \rho^4 }      
\qquad \qquad\qquad \qquad\qquad\quad
Z^2 = Z_1 Z_2-Z_4^2 \nn
\\
\beta_\varphi &=& \frac{ a^2 R  }{  \rho^2 }
\qquad\qquad\qquad\qquad\qquad\qquad\qquad\qquad\, 
\beta_\psi = 0 \nn \\
\gamma_\rho & =&  - { {\varepsilon}_1 R \over 2 \rho L_5^2} \, \Delta_n  \, \sin 2\phi
\qquad\,~ 
\gamma_\psi=\gamma_{\vartheta}=0   
\qquad\,\,\,\,	 
\gamma_v={\cal F}_n  
\\
\gamma_\varphi &=&\frac{ a^2 R  (1-{\cal F}_n) }{  \rho^2 }    -  {{\varepsilon}_1\,R\over 2 L_5^2 \rho^2}\,\Delta_n \cos 2\phi\,(\rho^2+a^2)  \nn  
\eea
%%%%%%%%%%%%%%%%%%%%%%%%%%%%%%%%%%%%%%%%%%%%%%%%%%%%%%%%%%%%%%%%%%%%%%%%%%%%%%%%%
Taking $\widehat P_{\vartheta}=  P_{\vartheta} = 0$, $\widehat{P}_\psi=P_{\psi} = 0$,  
the Hamiltonian reads
\be
\mathcal{H} = -P_u \widehat{P}_v + \frac{(\rho^2 + a^2)\,\widehat{P}_\rho^2}{2 Z^2 \rho^2} + \frac{\widehat{P}_\varphi^2}{2 Z^2 (\rho^2 + a^2)}
\ee
where the hatted conjugate momenta have the form
 \bea
 \widehat{P}_v  &=& P_\nu +{n\over R}\, P_\phi+ {\cal F}_n\, P_u      \nn\\
  \widehat{P}_\rho &=&  P_\rho +{{\varepsilon}_1 R P_u \, \Delta_n \sin 2\phi \over 2 \rho L_5^2}    \\
 \widehat{P}_\varphi &=&  P_\phi  -\frac{a^2}{\rho^2}\left(n\,P_\phi+R P_\nu + R P_u\right) +\frac{ 2 a^2   R\, P_u    }{   \rho^2   } \left[  {\cal F}_n + { {\varepsilon}_1  \Delta_n(\rho^2+a^2) \cos 2\phi\over  4 a^2 L_5^2}
  \right]   \nn
 \label{momentaplane12}
\eea
 with $P_u$ and $P_\nu$  conserved quantities.
 
 Let us focus on the truly dynamical variables $\rho$ and $\phi$. Their velocities are given by\footnote{The evolution of $\nu$, as well as of the other coordinates, follows from the one of $\rho$ and $\phi$.
 In particular   
 \be
\begin{aligned}
 \dot{\nu} = -P_u - \frac{a^2 R}{Z^2\,\rho^2(\rho^2 + a^2)}\widehat{P}_\varphi
\end{aligned}
\ee
 } 
 \be \label{velocities}
\begin{aligned}
\dot{\rho}&= \frac{\rho^2 + a^2}{Z^2\rho^2}\widehat{P}_\rho
\\
\dot{\phi} &= \frac{\widehat{P}_\varphi  (\rho^2-n a^2)}{Z^2\, \rho^2 \,(\rho^2 + a^2)}  -\frac{n P_u}{R} 
\end{aligned}
\ee
Choosing $\phi$ as independent variable, the equations of motion  can be written in the form
\bea
 {d\rho\over d \phi} &=& \frac{\widehat P_\rho \, R  (\rho^2+a^2)^2}{\widehat  P_\varphi \, R \, (\rho^2 -n  a^2) - P_u \, Z^2 \, \rho^2 \, (\rho^2 + a^2) }
\nn \\
{ d P_\phi \over d\phi}  &=& -{1\over \dot{\phi} }{\partial \over \partial \phi}   \left[\frac{(\rho^2 + a^2)\,\widehat{P}_\rho^2}{2 Z^2 \rho^2} + \frac{\widehat{P}_\varphi^2}{2 Z^2 (\rho^2 + a^2)}\right]  \label{eom}
\label{eomrp}
\eea
   and
   \be 
     \widehat P_\rho^2  =  {\rho^2\over (\rho^2+a^2)^2 } \left[    2 Z^2  \, P_u\, \widehat{P}_v  (\rho^2 + a^2)  -  \widehat P_\varphi^2     \right]  \label{pr2} \\
   \ee
We are interested in  solutions of the geodesic equations (\ref{eom}) characterised by trajectories trapped in the gravitational background. As before, we expect that for specific values of the 
incoming angular momentum $P_\phi$, there exists geodesics ending on trapping trajectories but now both the asymptotic trajectory and the angular momentum will in general vary with $\phi$.

 \subsubsection{Asymptotic  circular orbits}
 
   Due to the complexity of the three-charge problem along the $\vartheta=\pi/2$ plane, trajectories in general cannot be obtained in  analytic form. In this section we    
 present an example of solution in the region where the particle  reaches a critical orbit.  
     We  look for geodesics  asymptotically reaching circular  trajectories with constant  angular velocity, i.e.    $\dot{\rho}=0$, $\dot{\phi}=w$. For concreteness\footnote{We choose  $L_1=L_5=a$ only for illustrative purposes of the general case where the three quantities are of the same order. We notice that this symmetric choice is far different from the standard choice where $a$ is taken much smaller than the D1 and D5 charges, {\it i.e.} $a<< L_{1,5}$.}
 we take
\be
L_1=L_5=L=a    \qquad, \qquad      
\ee   
According to (\ref{velocities}), a  constant angular velocity  can be found by taking
    \be
   \rho^2 = n a^2  \qquad  \Rightarrow  \qquad    \dot{\phi}=-{n P_u\over R}  \label{ansatz}
\ee  
while  $\dot{\rho}=0$ requires
\be
  \widehat{P}_\rho=0  \label{solw}
\ee
or equivalently
\be
 2 Z^2  \, P_u\, \widehat{P}_v  (\rho^2 + a^2)  -  \widehat P_\varphi^2=0  \label{zeroh}
\ee
 We notice that at the critical point  $\rho^2 = n a^2$, $Z^2$ is constant and  $\widehat P_\varphi$ reduces to
\be
\widehat{P}_\varphi = \frac{R}{n}\left[2\mathcal{F}_n P_u - P_u - P_\nu + \frac{\varepsilon_1 (n+1)}{2\,a^2} P_u \Delta_n\cos 2\phi\right]
\ee 
  Equation (\ref{zeroh}) can therefore be easily solved for $P_\phi$
 \be
 P_\phi={R\over n} \left[   \frac{\widehat{P}_\varphi^2}{2 \,P_u \, Z^2\,a^2 (n+1)} -P_\nu -{\cal F}_n P_u \right]  \label{solp}
 \ee
The two equations of motion (\ref{eomrp}) are satisfied for $\rho = \sqrt{n}\,a$ and $P_\phi$ given by (\ref{solp}), quite remarkably this provides an exact solution for the non separable system.  It would be interesting to find a solution interpolating between infinity and these closed trajectories.

\subsection{ Geodesics in the near  horizon geometry}

 Finally, we consider  massless geodesics in the near horizon geometry. As shown in \cite{Bena:2017upb}, massless geodesics in this region are described by a separable dynamics that can be integrated in an analytic form.  
 The crucial difference with the case of asymptotically flat solutions is that in the near the horizon, $\phi$-oscillating terms are missing leading to solutions carrying no v-dependence. Here we display some simple examples of trapped geodesics in this region. The geodesics in this region can be viewed as the continuation of trajectories starting from infinity with initial conditions chosen such that no return or critical points are found before the particle reaches distances much smaller than $L_1$ and $L_5$.  
  
The near horizon geometry  is defined by taking  
 \be
 L_{1,5}^2 >> \rho^2+a^2      
 \ee
For this choice important simplifications take place. First, the regularity conditions (\ref{regularitycond}) reduce to
\be
\label{regularitycond}
\varepsilon_1=\varepsilon_4^2  \quad , \quad  \frac{L^2_1L^2_5}{ R^2}   = 2 a^2 + \varepsilon_4^2
\ee
with $L_1$, $L_5$, $R$ taken to be large with fixed ratio $L_1 L_5/R^2$.

The functions entering in the six-dimensional metric reduce to 
 \bea
Z^2 & =& \frac{\Delta_n s^2_\vartheta \left(2 a^2 R^2-L_1^2L_5^2\right)+L_1^2 L_5^2}{\left(a^2 c^2_\vartheta +\rho ^2\right)^2}\nn\\
\beta &=&\frac{ a^2 R \,  }{  \rho^2+a^2c_{\vartheta}^2}    \left[  s_{\vartheta}^2 \, d\varphi- c_{\vartheta}^2\, d\psi \right]  \nn\\  
\gamma &=&    \frac{ a^2 R \,    (1-{\cal F}_n)  }{  \rho^2+a^2c_{\vartheta}^2}    \left[  s_{\vartheta}^2 \, d\varphi+ c_{\vartheta}^2\, d\psi \right] +\mathcal{F}_n\, dv    \label{solutionads}
\eea
%%%%%%%%%%%%%%%%%%%%%%%%%%%%%%%%%%%%%%%%%%%%%%%%%%%%%%%%%%%%%%%%%%%%%%%%%%%%%%%%%
with  
\be
\begin{aligned}
  {\mathcal{F}}_n  &= -\frac{{\varepsilon}_4^2}{2a^2}\left[1-\left(\frac{\rho^{2}}{\rho^2+a^2}\right)^n\right] \\ 
 \Delta_n  &=  \frac{a^2}{\rho^2+a^2 }\left(\frac{\rho^2}{\rho^2+a^2}\right)^n         
\end{aligned}
\ee
 The Hamiltonian depends only on ${\vartheta}$ and $\rho$, so the momenta $P_u$, $P_v$, $P_\psi$ and $P_\varphi$ are all conserved. 
The Hamiltonian can be separated  according to   
\bea
 2 Z^2  \, (\rho^2+a^2 c_{\vartheta}^2) \,\mathcal{H} =\lambda_\rho (\rho,P_\rho) +\lambda_{\vartheta} ({\vartheta},P_{\vartheta})   \label{nulllambda}
\eea
with
\bea
\label{fuzzgeodsepKlambda}
\lambda_{{\vartheta}} ({\vartheta},P_{\vartheta})   &=&P_{{\vartheta}}^2 + {P_\psi^2 \over \cos^2{\vartheta}} + {P_{\varphi}^2 \over \sin^2 {\vartheta}}   \\
\lambda_{\rho} (\rho,P_\rho) &=& 
\left(a^2+\rho ^2\right) P_{\rho }^2+ \frac{2 R^2 P_u \left(2 a^2 \mathcal{F}_n+\epsilon _4^2\right) \left(\mathcal{F}_n P_u-P_v\right)}{a^2+\rho ^2} \nn\\
&& +\frac{a^2 \left( P_{\psi }+ R P_v-R P_u\right){}^2}{ \rho ^2} -\frac{a^2 \left( P_{\varphi }+R
   P_v + R P_u -2 \mathcal{F}_n R P_u\right){}^2}{a^2+\rho ^2}
\eea
\color{black}
 The equation $\mathcal{H} =0 $ can be solved by taking 
\be
\lambda_{{\vartheta}}=-  \lambda_{\rho} = K^2
\ee
with $K$ a constant, that can be interpreted as the total angular momentum.  Solving the second equation for $P_\rho(\rho)$ one finds
 \be
 P_\rho^2 (\rho)=  {{\cal P}_{2n+1}(\rho^2) \over  \rho^2 (   {\rho^2{+}a^2})^2  }  \label{p3c}
\ee
with ${\cal P}_{2n+1}(x)$ a polynomial of order $2n+1$. Turning points are associated to zeros of the polynomial ${\cal P}_{2n+1}(x)$ and critical geodesics to 
choices of angular momenta such that the two largest zeros of ${\cal P}_{2n+1}(x)$ collide.

For the sake of simplicity we will discuss only the $n = 1$ null geodesics, the order 3 polynomial reduces to
\begin{equation}
{\cal P}_{3}(x) = A x^3 + B x^2 + C x + D
\end{equation} 
where the list of coefficients reads
\begin{equation*}
\begin{aligned}
A &= - K^2
\\
B &=a^2 \left[\frac{2 \varepsilon_4^2 P_u P_v R^2}{a^2} +[P_\varphi + R (P_u+P_v)]^2-[P_\psi-R (P_u-P_v)]^2 - 2K^2\right]
\\
C &= a^4 \left[\left(\frac{\varepsilon_4^2 P_u R}{a^2}+[P_\varphi + R (P_u+P_v)]\right)^2 -2[P_\psi-R (P_u-P_v))]^2 - K^2\right]
\\
D &= -a^6\left[P_\psi -(P_u - P_v) R\right]^2
\end{aligned}
\end{equation*}

In order to illustrate the behaviour of the geodesics in this context, as before we choose
the conserved quantities such that $D = 0$, {\it i.e.} $P_\psi =R(P_u - P_v)$. A further simplification occurs by choosing $P_\varphi =- R(P_u + P_v)$
%For this choice the coefficients read
%\begin{equation*}
%\begin{aligned}
%A &= - K^2
%\\
%B &=-2a^2 K^2 + 4a^2 R^2P_u \left(P_u-P_v\right)+2 L_1^2 L_5^2 P_u P_v
%\\
%C &= -a^4 K^2 + P_u^2 \frac{ L_1^4 L_5^4}{R^2}
%\\
%D &= 0
%\end{aligned}
%\end{equation*}
%At last we can choose 
and $P_v = P_u$, leading to
\begin{equation}
{\cal P}_{3}(x) = -x  \left[ K^2 x^2 + 2\left(a^2 K^2 -\varepsilon_4^2 P_u^2 R^2\right) x + \left(a^4 K^2 - \varepsilon_4^4 P_u^2 R^2\right)\right]
\end{equation}
by requiring two coincident roots one finds the relations
\begin{equation}
\rho_\text{crit} = \sqrt{a^2 - \varepsilon_4^2}
\qquad
,
\qquad
K^2 = \frac{\varepsilon_4^2 P_u^2 R^2}{2 a^2 - \varepsilon_4^2}
\end{equation}
This shows that critical geodesics exist if $a > \varepsilon_4$ {\it i.e.} $aR>L_1L_5/\sqrt{3}$.

%For example, taking   
%\be
%n=1 \quad , \quad   P_\psi=P_u-P_v=P_\varphi+2 R P_v=0 \quad , \qquad K^2={\epsilon_4^2\, R^2\, P_v^2 \over  2 a^2-\epsilon_4^2}    
%\ee
%one finds
%\be
%{\cal P}_3(x)= K^2\, x^3+2 \, x^2 (a^2\, K^2+\epsilon_4^2 \, R^2\, P_v^2)+x\, (a^4\, K^2+\epsilon_4^4 \, R^2\, P_v^2)   
%\ee
%A critical trajectory is found for
%\be
%K^2={\epsilon_4^2 \, R^2\, P_v^2\over 2 a^2-\epsilon_4^2} 
%\ee
% approaching asymptotically a  limiting  trajectory at radius 
%\be
%\rho_*=\sqrt{a^2-\epsilon_4^2}
%\ee

\section{Conclusions and outlook}
 \label{outlook} 

Relying on a class of micro-state geometries for 3-charge systems in $D=5$ constructed in \cite{Bena:2017xbt}, we have further tested the fuzzball proposal by studying massless geodesics in these backgrounds. 
In particular we have shown that 2- and 3-charge fuzzball geometries tend to trap massless neutral particles  for a specific choice of their  impact parameter.  This is at variant with classical BH's that trap all particles impinging with an impact parameter below a certain critical value of the order of the horizon radius.  This suggests that the blackness  property of black holes arises as a collective effect whereby each micro-state absorbs a specific channel.  

The analysis has been performed in various steps. First we have reviewed the general form of the metric and written down the geodesic equations for massless neutral probes in both the Lagrangian and Hamiltonian forms. Then we focused on the cases of (singular) non-rotating BPS black-holes with 3-charge, on micro-states for 2-charge systems with a circular profile and finally on the 3-charge case.

We have (implicitly) integrated the geodesic equations for the 2-charge case for generic initial values of the angle $\vartheta$ and of the integration constant $K$ (playing the role of total angular momentum), thus generalising our previous results for ${\vartheta} = 0$ (plane orthogonal to the circular profile) and  ${\vartheta} = \pi/2$ (plane of the circular profile). 

In the 3-charge case we have fully analysed the geodesics for ${\vartheta} = 0$  (since they lead to separable equations of the same form as in the 2-charge case, previously analysed) and written down the equations for  ${\vartheta} = \pi/2$, that lead to a non-separable  system. A simple solution of this intricate system has been found.  

We also considered massless geodesics on asymptotically AdS 3-charge geometries of the type studied in \cite{Bena:2017upb}\footnote{We thank the referee for drawing our attention on this work.}. 
These geometries, unlike their extension to asymptotically flat space, are characterized by a separable dynamics and massless geodesics can therefore be integrated in an analytic form. We presented explicit examples of trapped geodesics that can be viewed as the end points of the trajectories of massless particles infalling from infinity without encountering turning or critical points before reaching distances much smaller than $L_1$ and $L_5$.

In this paper we restricted our attention to the study of scattering of classical point-like massless neutral probes. It would be interesting to extend this analysis to more general probes like massive, possibly charged, particles, waves and strings where tidal effects such as those studied in
\cite{Tyukov:2017uig} can be relevant.

Other classes of smooth (non-supersymmetric) geometries (such as JMaRT \cite{Jejjala:2005yu})  lead to interesting effects \cite{Bianchi:wip} due to the presence of an ergo-region of finite extent without horizons or singularities. In \cite{Eperon:2016cdd}, the authors studied the properties of geodesics in the closely related setup of five and six dimensional supersymmetric fuzzball geometries.  In particular they used the presence of stably trapped geodesics to argue for the existence of a non-linear instability even for these BPS microstate geometries. 
These trapped geodesics may be related to the circular orbits considered in section 5.3.1 of the present paper. It would be interesting to study  linear perturbations and (quasi-)normal modes that may signal a potential instability of the microstate solutions. 

Finally, the analysis in \cite{Frolov:2017kze} has some overlap with section 3 of the present paper, where for completeness and comparison with the original results of our analysis we discussed null geodesics in rotating and non-rotating singular black-holes in five dimensions.

 \section*{Acknowledgements}
  We acknowledge fruitful discussions with Andrea Addazi, Pascal Anastasopoulos, Guillaume Bossard, Ramy Brustein, Paolo Di Vecchia, Maurizio Firrotta, Francesco Fucito, Stefano Giusto, Elias Kiritsis, Antonino Marcian\`o, Lorenzo Pieri, Gabriele Rizzo, Rodolfo Russo, Raffaele Savelli, Masaki Shigemori, Gabriele Veneziano, and Natale Zinnato. 
Part of the work was carried on while D.~C. was visiting Erwin Schr\"odingier Institute in Vienna and while and M.~B. and J.~F.~M. were visiting Galileo Galilei Institute in Florence. We would like to thank both Institutes for the kind hospitality. We thank the MIUR-PRIN contract 2015MP2CX4002 {\it ``Non-perturbative aspects of gauge theories and strings''} for partial support.

 \clearpage

\bibliographystyle{JHEP}
\bibliography{references_DarkSideFuzzball}
 
\end{document}